\RequirePackage{fix-cm}
\documentclass[smallextended,final]{svjour3}       

\usepackage{graphicx}
\usepackage{mathptmx}      
\usepackage{amsmath}

 \journalname{Exp Astron}

\begin{document}

\title{High Contrast Observations of Bright Stars with a Starshade}
\author{Anthony Harness$^{1,3}$ \and Webster Cash$^1$ \and Steve Warwick$^2$}
\institute{A. Harness \at anthony.harness@princeton.edu\\
	\and $^1$ Dept. of Astrophysical \& Planetary Sciences, University of Colorado Boulder, Boulder, CO
	\at $^2$ Northrop Grumman Aerospace Systems, Redondo Beach, CA
	\at $^3$ Dept. of Mechanical \& Aerospace Engineering, Princeton University, Princeton, NJ
}
\date{Received: 15 June 2017 / Accepted: 24 November 2017}

\maketitle

\begin{abstract}
Starshades are a leading technology to enable the direct detection and spectroscopic characterization of Earth-like exoplanets. In an effort to advance starshade technology through system level demonstrations, the McMath-Pierce Solar Telescope was adapted to enable the suppression of astronomical sources with a starshade. The long baselines achievable with the heliostat provide measurements of starshade performance at a flight-like Fresnel number and resolution, aspects critical to the validation of optical models. The heliostat has provided the opportunity to perform the first astronomical observations with a starshade and has made science accessible in a unique parameter space, high contrast at moderate inner working angles. On-sky images are valuable for developing the experience and tools needed to extract science results from future starshade observations. We report on high contrast observations of nearby stars provided by a starshade. We achieve $5.6\times10^{-7}$ contrast at 30 arcseconds inner working angle on the star Vega and provide new photometric constraints on background stars near Vega.

\end{abstract}

\section{Introduction}
\label{sec:introduction}
Starshades are a leading high contrast technology to enable the direct detection and spectroscopic characterization of Earth-like exoplanets \cite{Cash_2006}. Their high throughput and inherent broadband coverage bring the technically challenging goal of spectroscopy of extremely faint exoplanets into the realm of possibility. Their ability to achieve high contrast at an inner working angle (IWA) that is decoupled from the telescope size and without the use of active wavefront control provides the opportunity for most future observatories to be readily adapted to work with starshades. The simplicity of the telescope behind the starshade also means that science with starshades can begin now and at low cost.

A strong motivation for a low cost observing program with starshades is to build experience with and develop analysis tools for the data products of a future starshade mission. The novelty of the starshade architecture means we must start with the basics in understanding how the system operates as a whole and how observations lead to final science products. While internal coronagraphs draw from decades of experience in observations and testing, no such heritage exists for starshades and we must start from square one with system level demonstrations. We believe an effective strategy to developing the starshade experience base is with real astronomical observations and data products similar to what is expected from a full-scale mission. 

As with any new scientific experiment that is probing orders of magnitude deeper than before, there will be technical surprises. And space experiments are so very expensive we cannot afford to be blind-sided by ``unknown unknowns''. Testing of scale model starshades in the lab have shown that the basic concept of the starshade is sound \cite{Leviton_2007}, but system design and operation issues remain to be demonstrated at a level adequate for mission risk mitigation. Some of the potential system issues that have been identified are 1) alignment of starshades along the line of sight 2) maintenance of that formation across tens of thousands of kilometers in space autonomously 3) scattering of starlight from behind the starshade 4) scattering of sunlight off the edges of the starshade 5) modeling of residual starlight and background near the edge of the starshade and 6) calibration of images and spectra near the edge of starshade. Remarkably, starshades function well in air, subject mostly to the limitations of ``seeing''. So starshades can and must be tested in air to learn as much as possible quickly and at minimal cost.

Working with sub-meter starshades at a few kilometers separation provides an opportunity for science in a unique region of parameter space, high contrast but at a moderate IWA. Most high contrast instruments currently in use on large ground-based telescopes have a limited field of view of no more than a few arcseconds. These major facilities are also expensive to use and competitive in allocated time. The starshade-heliostat facility detailed in this paper, using the McMath-Pierce Solar Telescope (hereafter McMath) at the Kitt Peak National Observatory, has the potential to fill a unique niche of science. It has the potential to achieve high contrast at IWAs inaccessible to current state-of-the-art high contrast instruments. Additionally, the availability and low operational cost of McMath make it an appealing option for experimenting with starshades. Our efforts at McMath are designed to learn as much as we can about the system level realities of starshades without pushing on the limits of IWA and contrast that will have to be achieved eventually.

Here, we report on high contrast observations of nearby stars provided by a starshade. A 24 cm diameter starshade at a separation of 2.4 km allows us to reach an IWA of 10 arcseconds while operating at a flight-like Fresnel number of 12. In this configuration, we achieve a contrast of $2.0\times10^{-5}$. Operating at a shorter baseline with a relaxed Fresnel number of 22, we achieve $5.6\times10^{-7}$ contrast at 30 arcseconds IWA on the star Vega and provide new photometric constraints on background stars near Vega. These observations use a starshade in the most flight-like configuration (in terms of Fresnel number, IWA, and telescope resolution) to date and provide a key demonstration for advancing the technology of starshades. 

This paper is organized as follows. Section~\ref{sec:experiment} gives a brief summary of the experimental setup using McMath, further details of which can be found in Refs.~\cite{Harness_2016}, \cite{Harness_thesis}. Section~\ref{sec:observations} summarizes the observations made and Section~\ref{sec:analysis} details the data analysis procedures. Section~\ref{sec:results} comprises the bulk of the paper and reports the main results. Section~\ref{sec:discussion} gives a brief discussion on what was learned from this first demonstration of a new facility and Section~\ref{sec:improvements} suggests improvements to be made in future tests. Finally, Section~\ref{sec:conclusions} summarizes and concludes.

\section{Experiment Description}
\label{sec:experiment}
The distributed architecture of the starshade and the necessity to maintain alignment between telescope, starshade, and star presents a challenge in testing the system at the correct scaling of starshade size and separation. The scaling is driven by the motivation to observe at angles close to the host star where more scientifically interesting targets lie, while maintaining a large enough Fresnel number to provide sufficient contrast. The Fresnel number,
\begin{equation}
	N_F = \frac{a^2}{\lambda F}
	\label{eq:fresnel_number}
\end{equation}
\noindent with starshade radius $(a)$, separation $(F)$, and wavelength $(\lambda)$, is a similarity parameter that, when held constant, enables us to test the physics of a full-scale starshade mission on much smaller scales. Therefore, as long as the Fresnel number is conserved, sub-scale demonstrations on the ground can be done to prepare for a future space mission. Previous tests of starshades have been done indoors \cite{Leviton_2007,Samuele_2009,Cady_2010,Sirbu_2016,Kim_2016} and on extremely flat dry lake beds \cite{Glassman_2014}, but at Fresnel numbers that were significantly larger than the flight-like Fresnel number ($\sim14$) and at large inner working angles. Current lab tests are pushing towards flight-like Fresnel numbers, but the smaller separations require the use of very small starshades where microscopic features become difficult to manufacture \cite{Kim_2017}. The large separations provided by McMath allow us to test macroscopic starshades at the correct Fresnel number and allow us to push down in IWA. 

A further limitation of previous tests was that they used artificial light sources and had no clear path of extending the tests to work with astronomical targets. The motion of astronomical objects across the night sky adds a significant layer of complexity to maintaining alignment. Previous efforts for astronomical observations attempted to use suborbital vehicles such as Zeppelins \cite{Harness_2013} and vertical takeoff vertical landing rockets \cite{Sorgenfrei_2017} to maintain alignment, but were limited by the flight stability and achievable altitude of the vehicles.

\begin{figure}
	\includegraphics[width=\textwidth]{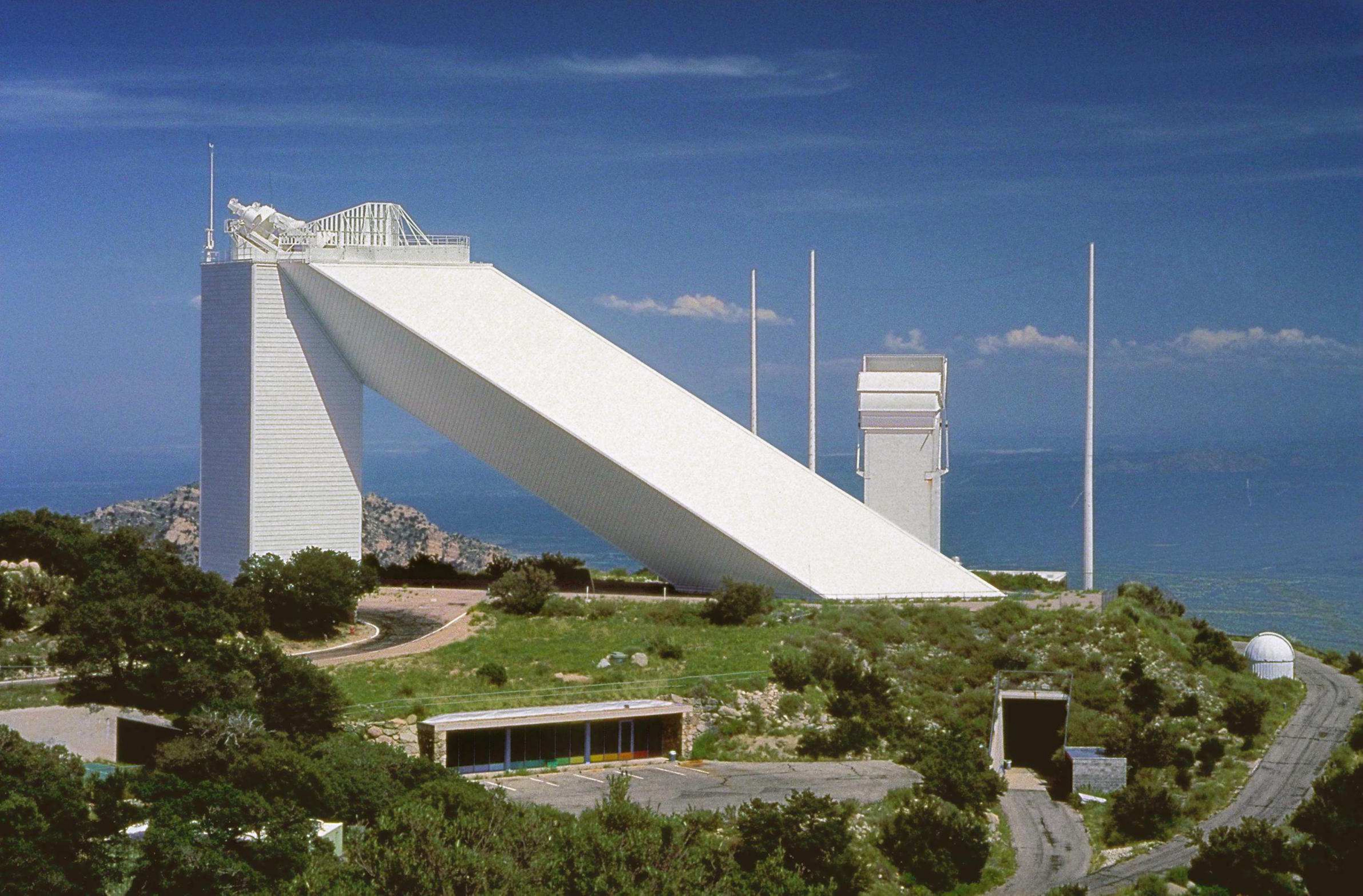}
	\caption{McMath-Pierce Solar Telescope at the Kitt Peak National Observatory. Image credit: KPNO.}
	\label{fig:mcmath}      
\end{figure}
\begin{figure}
	\includegraphics[width=\textwidth]{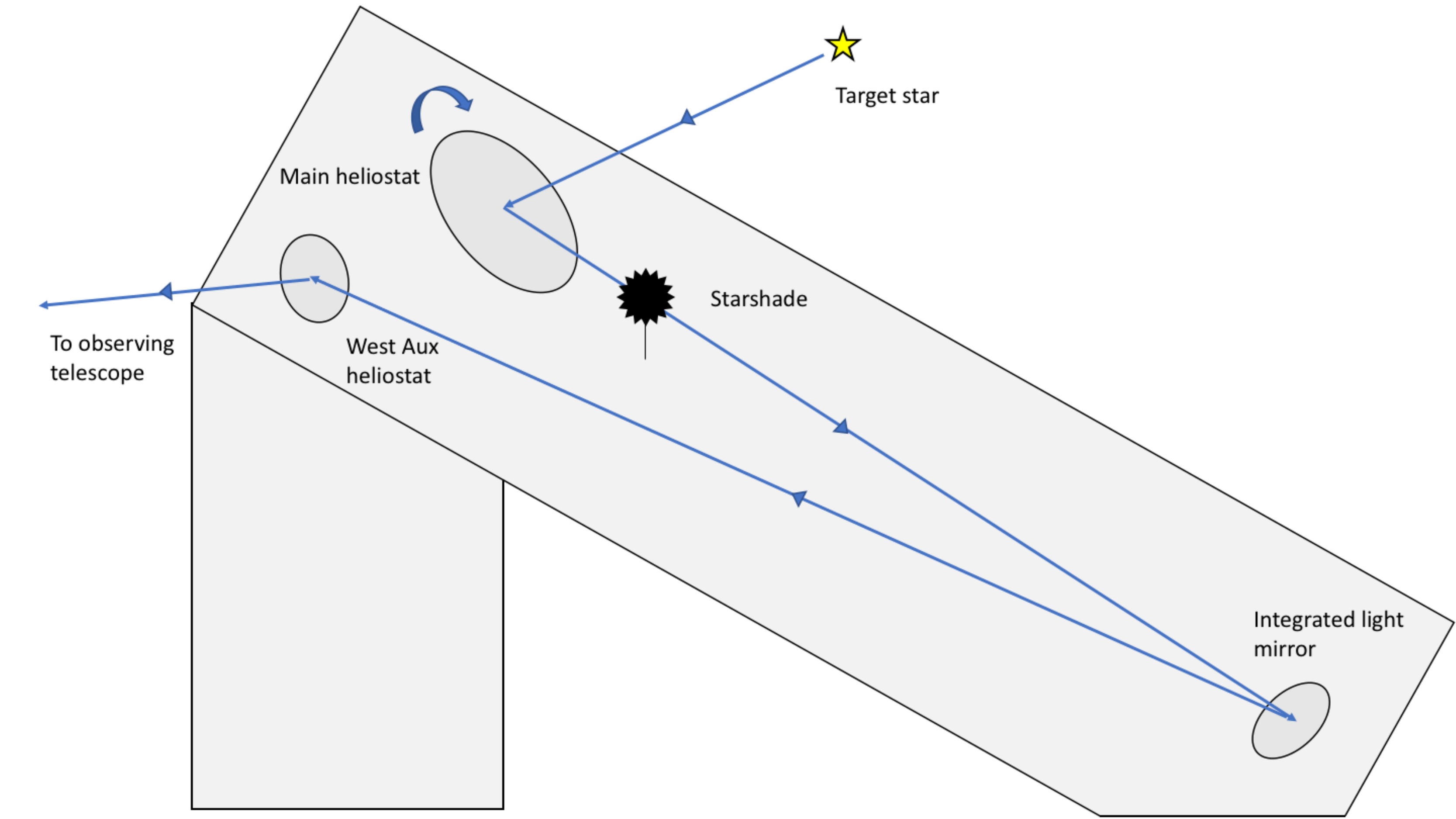}
	\caption{Schematic of the optical layout of the McMath facility. Incoming starlight bounces off 3 flat mirrors before being sent out of the facility towards our observing telescope.}
	\label{fig:optical_layout}      
\end{figure}

In early 2015, a solution to the alignment problem was found with the McMath-Pierce Solar Telescope. The main heliostat of the solar telescope can be used to track stars across the sky and provide a stable beam of parallel light incident on a sub-scale starshade, which can then be observed by a small telescope at a distance. Initial tests and the first astronomical observations were performed with a 400 m baseline between the starshade and telescope \cite{Novicki_2016}. More recently, a closed loop tracking system for the heliostat was developed to correct for misalignments due to tracking errors and atmospheric turbulence \cite{Harness_2016}. This allowed the baseline to be extended to 2.4 km, enabling observations at a smaller inner working angle while maintaining a flight-like Fresnel number and resolution. More details on the experimental setup and the closed-loop tracking system can be found in Ref.~\cite{Harness_2016} and Ref.~\cite{Harness_thesis}. 

The optical layout of the McMath system is shown schematically in Figure~\ref{fig:optical_layout}, with parameter values listed in Table~\ref{tbl:starshade_parameters}. Incoming starlight hits the main 2 m diameter heliostat and is sent down the tunnel, where it immediately hits a starshade supported on a stand. Halfway down the tunnel, a 1 m flat mirror intercepts the light and sends it back up the tunnel to the 0.9 m west auxiliary heliostat flat mirror, located slightly offset from the main heliostat. The west auxiliary directs the light beam out of the facility and towards the direction of our observing telescope located on a far ridge of Kitt Peak. All mirrors in the system are flat and fixed, except for the main heliostat, which is rotating with the celestial sphere.

\begin{table}
\caption{Optical parameters of the experiment in the {\bf short} and {\bf long} baseline configurations. The starshade diameter, inner working angle, and Fresnel number are referenced to the $1/e$ inflection point of the starshade.}
\label{tbl:starshade_parameters}     
\begin{center}
\begin{tabular}{cccc}
	\hline\noalign{\smallskip}
        {\bf Parameter} & {\bf Short Baseline} & {\bf Long Baseline} \\ 
        	\noalign{\smallskip}\hline\noalign{\smallskip}
        Starshade-telescope separation: & 570 m & 2.4 km \\ 
	Starshade diameter: & 16 cm & 24 cm\\ 
        Inner working angle: & 30'' & 10''\\ 
        Fresnel number:  & 22 & 12\\ 
        Telescope diameter: & 2 cm & 2 cm\\ 
        Resolution elements across starshade: & 11 & 4\\ 
        Wavelength coverage: & 400 nm - 800 nm & 400 nm - 800 nm\\ 
	\noalign{\smallskip}\hline	
\end{tabular}
\end{center}
\end{table}

\begin{figure}
	\includegraphics[width=\textwidth]{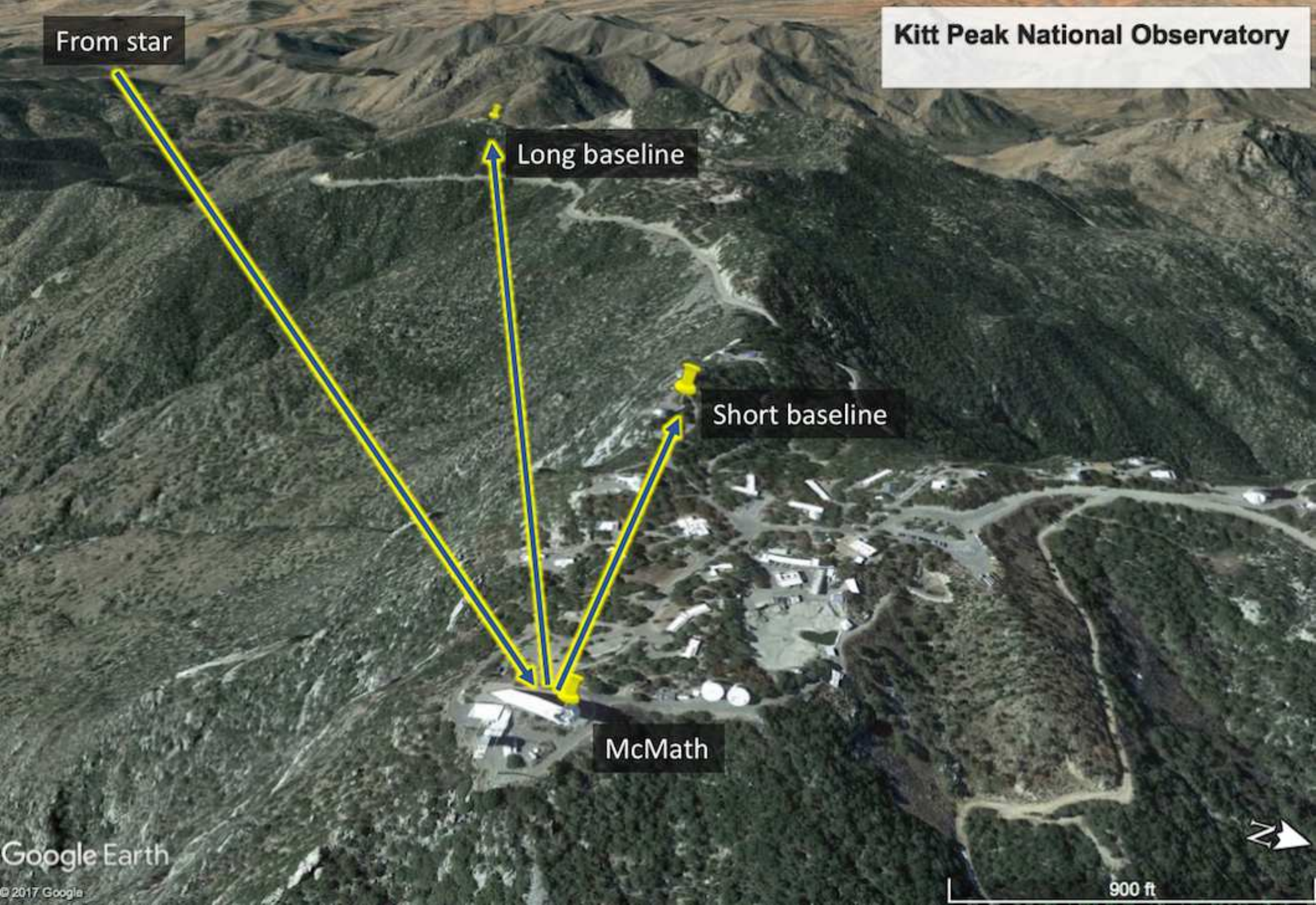}
	\caption{Google Earth map of Kitt Peak. Starlight is sent from McMath to the observing telescope at either the short or long baseline positions.}
	\label{fig:google_earth}      
\end{figure}

The observing telescope is located away from McMath at either the parking lot of the 0.9 m telescope (short baseline) or the SW Ridge (long baseline). Figure~\ref{fig:google_earth} shows the locations of the short and long baselines on the mountain. The observing telescope is a 13 cm refractor from Explore Scientific that is stopped down to a 2 cm - 6 cm aperture. A 2.5x Barlow lens is used to increase the focal length to $\sim$2.5 m. The detector behind this telescope is an Andor iXon 897 Electron Multiplying CCD (EMCCD). EMCCDs differ from traditional CCDs in that they have a series of multiplying stages in the readout register that provide gains of up to 1000x. This gain effectively eliminates any readout noise and allows for short exposures at a high readout rate without sacrificing signal-to-noise performance. Using this detector allows us to continuously take many short ($\sim0.1$ s) exposures and keep the exposures in which we have good alignment behind the starshade and throw out those in which light leaking around the starshade is contaminating the signal.

\begin{figure}
	\includegraphics[width=\textwidth]{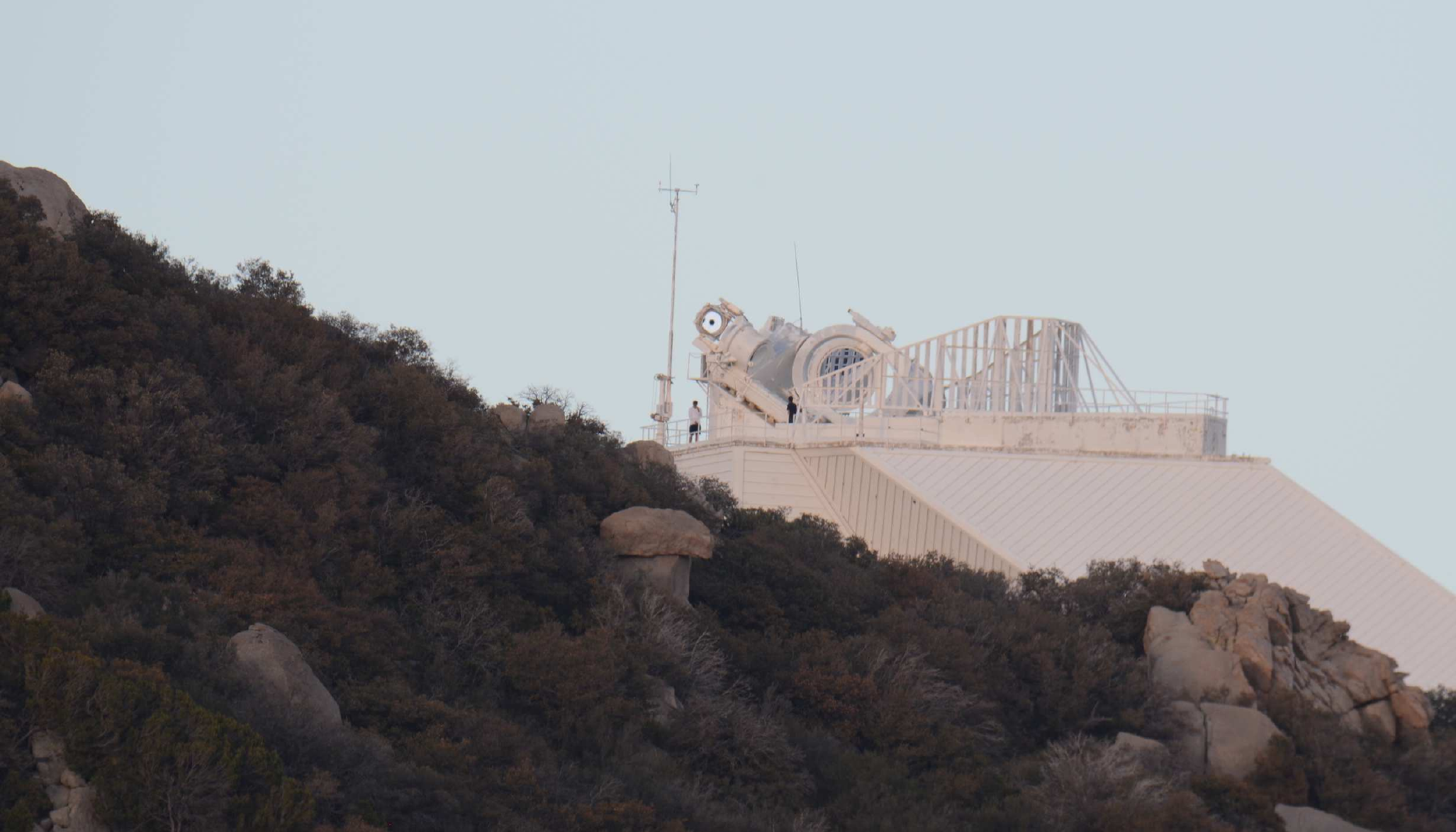}
	\caption{View from observing telescope on SW Ridge. The dark starshade can be seen in the center of the west auxiliary mirror.}
	\label{fig:mcmath_view}      
\end{figure}

\section{Observations}
\label{sec:observations}
We conducted two full observing runs at McMath, the first in November 2015, the second in April 2016. A shorter run was completed in early July 2016 to experiment on the mirrors in the system and test at the short baseline. The dates of these runs, along with the observed targets, are shown in Table~\ref{tbl:observations}. 
\begin{table}
\caption{Summary of observations with the total exposure time for each target. The total exposure time does not necessarily reflect the amount of usable data.}
\label{tbl:observations} 
\begin{center}    
\begin{tabular}{cllcc}
	\hline\noalign{\smallskip}
	{\bf Run} & {\bf Dates} & {\bf Notes} & {\bf Target} & {\bf Total Exposure Time [s]} \\
	\noalign{\smallskip}\hline\noalign{\smallskip}
	1 & 11/9/15 -  & -- First test from long baseline & Fomalhaut & 8400 \\
	 & 11/13/15& -- No feedback loop & Aldebaran & 670 \\
	 & & & Vega & 290 \\ 
 	\noalign{\smallskip}\hline\noalign{\smallskip}
	 2 & 4/5/16 -  & -- First test of feedback loop & Rasalgethi & 3400\\
	 & 4/9/16& & Arcturus & 2900 \\
	 & & & Sirius & 1800 \\
	 & & & Altair & 1200 \\
	 & & & Vega & 760 \\ 
 	\noalign{\smallskip}\hline\noalign{\smallskip}
	 3 & 7/11/16 - & -- Short baseline observations & Vega & 430 \\
	 &  7/13/16&& Antares & 360 \\
	\noalign{\smallskip}\hline	
\end{tabular}
\end{center}
\end{table}

For each target observed, the heliostat must be aligned to place the target in the field of view of the observing telescope, whose view is limited by the angular extent of the west auxiliary mirror as viewed from a distance. The view of the west auxiliary mirror as seen from the long baseline is shown in Figure~\ref{fig:mcmath_view}. The alignment procedure (detailed further in Ref.~\cite{Harness_thesis}) is a trial-and-error process that has a large overhead time, so we try to minimize the number of target changes each night. After the target is in view of the observing telescope, the feedback loop is calibrated by observing the motion of the star while the heliostat is stepped along orthogonal axes. 

A typical observation consists of staying on one target and continuously taking images of the star behind the starshade. There is small management required to keep the tracking system updated to account for the tracking telescope and science telescope sampling different patches of atmosphere. We observe all images as they are downloaded in real time to assess the atmospheric conditions and change the exposure time accordingly. Once or twice an hour, the target is moved from behind the starshade and images of the unblocked star are taken to provide a photometric calibration. Ideally, the unblocked images are taken frequently and close in time to the blocked images to sample the same atmospheric conditions, but the large overhead time in realignment places a practical limitation on how frequently this is done.

\begin{figure}
	\includegraphics[width=\textwidth]{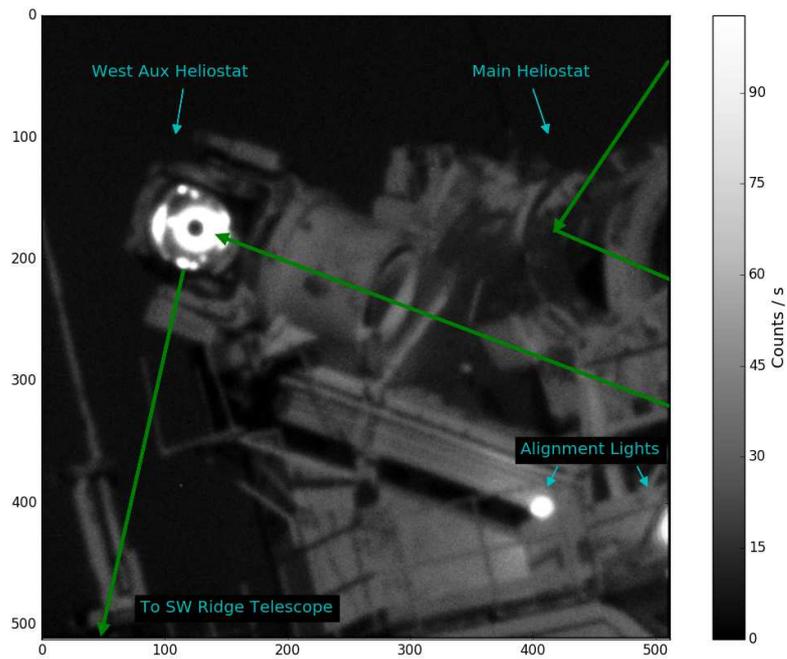}
	\caption{View from observing telescope on SW Ridge. The dark starshade can be seen in the center of the west auxiliary mirror, surrounded by diffracted light. The alignment lights are in the lower right corner. The path of the light beam is shown in green.}
	\label{fig:mcmath_image}      
\end{figure}

\section{Data Analysis}
\label{sec:analysis}
The data product from the observations comprise a large number ($\sim$30,000 per night) of short exposure EMCCD images. The image processing of such a large dataset, particularly low signal images in a varying atmosphere and on an unstable platform, is not trivial. There are a lot of moving parts in this optical system that makes image registration difficult. The observed star field is an image projected by the heliostat mirror, which is moving in a random motion as the tracking system is compensating for misalignments. This is also complicated by the low resolution resultant of the large pixel size of the camera and small aperture. However, image stacking is necessary as the star is continuously moving in and out of alignment. Using the EMCCD to take many short exposure images allows us to regain some signal-to-noise by preferentially choosing the best images to be combined.
 
\subsection{Image Alignment}
\label{sec:alignment}
For a given night, the position of the starshade is fixed in the image, while the star field rotates with the celestial sphere. Vibrations in the observing telescope, primarily due to wind buffeting, cause the entire scene (the starshade and star field) to shift on the detector between exposures. We correct for this vibrational jitter by registering each image on two alignment lights mounted to the exterior of McMath (see Figure~\ref{fig:mcmath_image}) that are visible on the detector for every exposure. Since the image motion is due to vibrational jitter in the telescope, correcting for the jitter in the alignment lights will also correct for the jitter in the star field. For each image, the positions of the alignment lights are found with an intensity-weighted centroid and all images observed on the same night are translated to register on the lights. Image motion due to atmospheric disturbances will be different between the alignment lights and the starshade, however, we do not attempt to correct for this and assume the motion is dominated by seeing effects on a much shorter timescale.

On a night-to-night basis, the starshade is not necessarily in the same position as the west auxiliary heliostat does not unstow to the exact position. Therefore images between nights must also be aligned to a common center.  For the short baseline, there are a few background stars in the field of view and the images between nights are registered on the star field. For the long baseline, the field of view through the heliostat is only 90 arcseconds across and there are no background stars to align with, so we register the images between nights on the starshade.

\subsection{Unblocked Target Intensity}
\label{sec:unblocked}
To calculate how much light was suppressed by the starshade and to provide photometric calibrations of background stars, we need to know the intensity of the target star without the starshade in place. However, the intensity of the unblocked source is highly variable due to the light traveling along a large horizontal path of atmosphere. The atmosphere affects the measured intensity in two ways, the point spread function (PSF) of the telescope is broadened and light is spread over a larger area on the detector, and scintillation and atmospheric extinction cause a random fluctuation in the amount of light entering the telescope. We will combine the effects of scintillation and changes in atmospheric extinction, i.e., variations in the intensity, into the single term of ``scintillation''. Without knowledge of how quickly the atmospheric extinction is changing, it is not possible to distinguish the two.

\begin{figure}[h]
	\includegraphics[width=\textwidth]{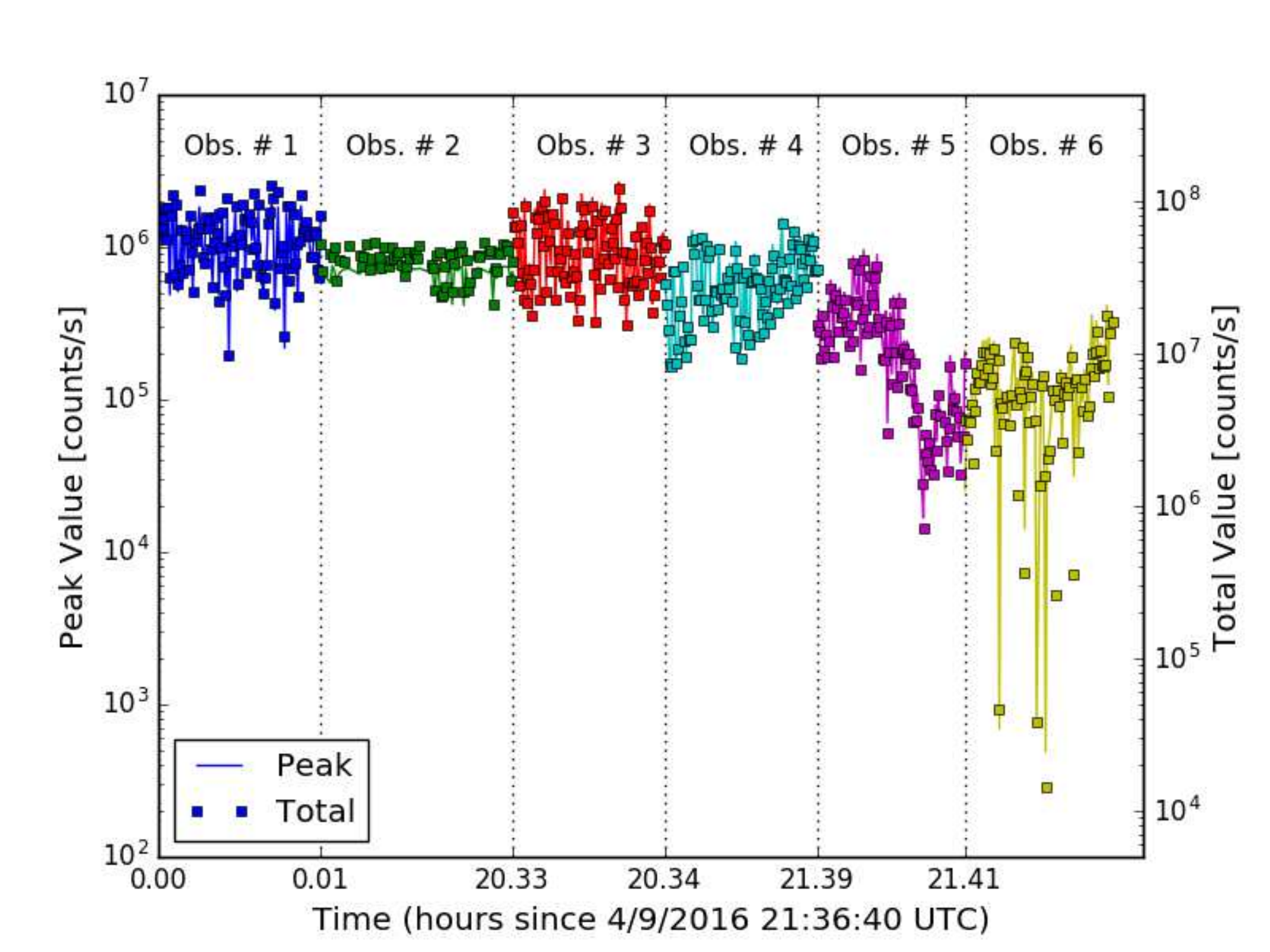}
	\caption{The intensity of Arcturus, peak value ({\bf line}) and total ({\bf squares}), as a function of time for unblocked images of Arcturus over the course of two nights. These data are taken from the {\bf long} baseline. The vertical lines mark different groups of exposures. The variance in the intensity is due to scintillation in the atmosphere. The lower values in Observations \#5 \& \#6 are likely due to sampling different atmospheric properties.}
\label{fig:unblocked_time}
\end{figure}

Scintillation is by far the largest source of error in our photometric estimations and the large uncertainty in the measured intensity of the target star is propagated to a large uncertainty in the performance of the starshade, making results difficult to interpret. To examine the effect the atmosphere has on our estimate of the unblocked target intensity, we split the unblocked images of Arcturus into six distinct 30 second observations over three nights. Figure~\ref{fig:unblocked_time} shows the peak intensity and total energy (counts summed over a subaperture in the image) of the unblocked source as a function of time. These values and their errors are tabulated in Table~\ref{tbl:unblocked_values}. There is a clear difference in the target intensity between observing nights as we are observing in different atmospheric conditions.

The relative contributions from the other dominant error terms are provided in Table~\ref{tbl:unblocked_errors}. $\sigma_\text{observed}$ is the observed variation in the intensity, $\sigma_\text{shot}$ is the photon noise following Poisson statistics, $\sigma_\text{gaussfit}$ is the RMS error in fitting a gaussian to find the peak intensity, and $\sigma_\text{scintillation}$ is the expected variance from scintillation. The scintillation variance is calculated from the scintillation index, $\sigma^2_I$, which is given by,
\begin{equation}
	\sigma^2_I = \frac{\left<I^2\right>  - \left<I\right>^2}{\left<I\right>^2}
\end{equation}
\noindent where $\left<I\right>$ is the mean intensity.

\begin{figure}[h]
	\begin{center}
		\includegraphics[width=\textwidth]{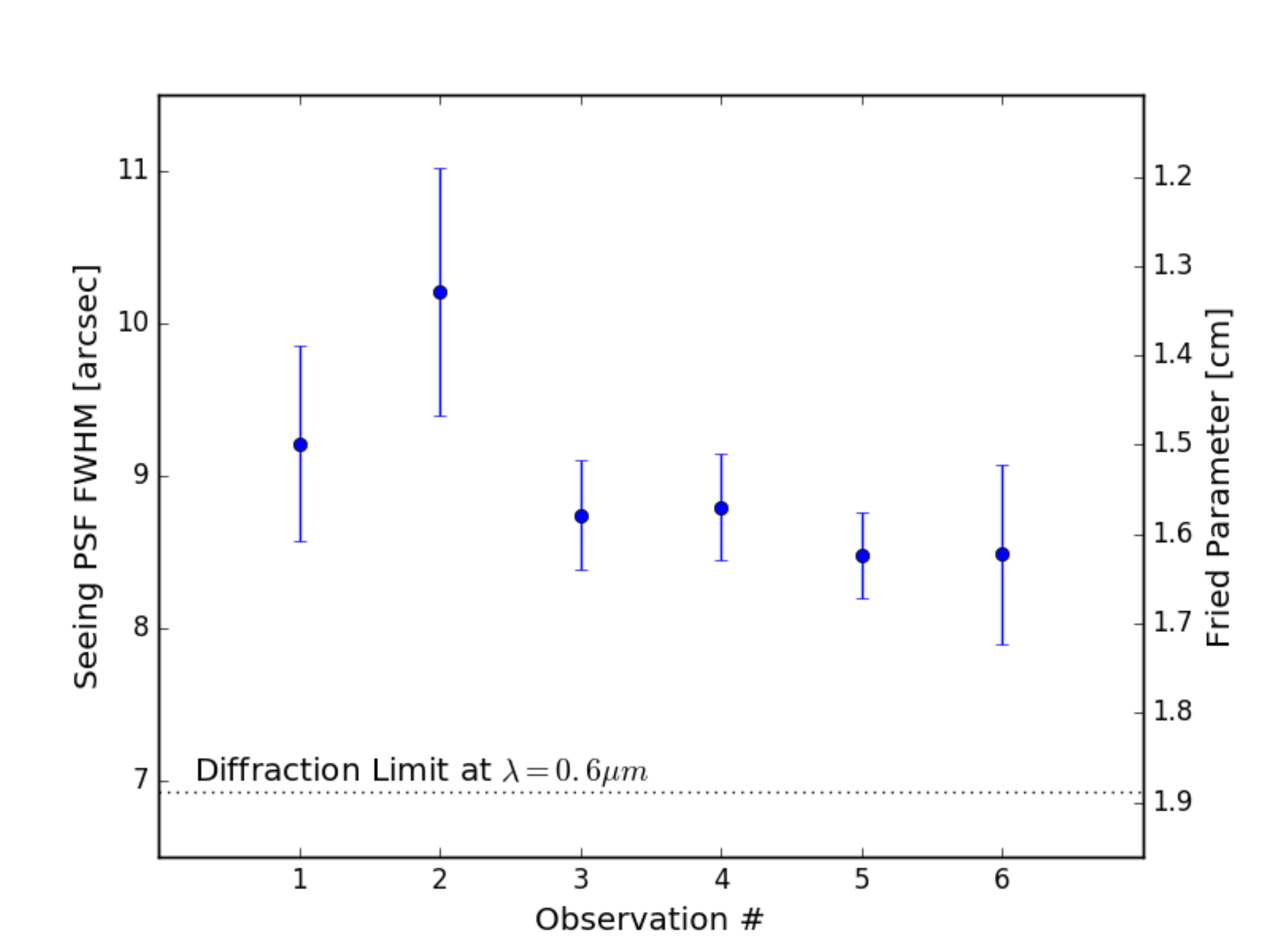}
	\end{center}
	\caption{The estimate of atmospheric seeing, obtained from the width of the PSF in the image, for the different observations of Figure~\ref{fig:unblocked_time}. The large seeing suggests very strong atmospheric turbulence.}
\label{fig:r0}
\end{figure}

Table~\ref{tbl:unblocked_errors} shows that nearly all the error can be attributed to scintillation. The exception being Observation \#5 in which a large systematic decrease in the intensity is seen, possibly due to a stable change in the atmospheric properties during those exposures. The exposure time in Observation \#2 is longer than the others and looks more stable, but it has an upper cutoff and is believed to be saturated. As intensity variations dominate over the PSF broadening, we find there is little difference in using the peak intensity vs the total energy in the PSF to calculate the unblocked value.
\begin{table}
	\caption{The mean ($\mu$), standard deviation ($\sigma$), and percent error of the peak count rates for observations of unblocked Arcturus from the {\bf long} baseline. The observation numbers correspond to those in Figure~\ref{fig:unblocked_time}. At times, the error in the intensity of the unblocked source is up to 100\%.}
	\label{tbl:unblocked_values}
\begin{center}
        \begin{tabular}{ccccccc} 
	\hline\noalign{\smallskip}
        {\bf Observation} & {\bf 1} & {\bf 2} & {\bf 3} & {\bf 4} & {\bf 5} &  {\bf 6} \\ 
	\noalign{\smallskip}\hline\noalign{\smallskip}
       $\mu~(\times 10^{5}$ ct/s) & 11.9 & 7.0 & 9.4 & 6.0 & 2.3 & 1.3\\ 
	\noalign{\smallskip}\hline\noalign{\smallskip}
       $\sigma~(\times 10^{5}$ ct/s)  & 4.64 & 0.83 & 5.40 & 3.12 & 2.18 & 0.77 \\ 
	\noalign{\smallskip}\hline\noalign{\smallskip}
       \% Error  & 39.0 & 11.9 & 57.2 & 51.9 & 95.2 & 58.3 \\ 
        \noalign{\smallskip}\hline	
        \end{tabular}
\end{center}
\end{table}
\begin{table}
\caption{Relative contribution (percent error) of different sources of error to the total error in the {\bf unblocked} source intensity. The variance in the intensity due to scintillation in the atmosphere is by far the dominant error term. These data are for Arcturus from the {\bf long} baseline.}
\label{tbl:unblocked_errors}
\begin{center}
\begin{tabular}{lcccccc} 
	\hline\noalign{\smallskip}
        {\bf Observation} & {\bf 1} & {\bf 2} & {\bf 3} & {\bf 4} & {\bf 5} &  {\bf 6} \\ 
	\noalign{\smallskip}\hline\noalign{\smallskip}
       $\sigma_\text{observed}$ & 39.0 & 11.9 & 57.2 & 51.9 & 95.2 & 58.3\\ 
	\noalign{\smallskip}\hline\noalign{\smallskip}
       $\sigma_\text{scintillation}$  & 38.9 & 12.3 & 49.8 & 48.8 & 76.0 & 59.0 \\
	\noalign{\smallskip}\hline\noalign{\smallskip}
       $\sigma_\text{gaussfit}$  & 2.2 & 4.5 & 2.2 & 2.4 & 3.6 & 4.6 \\ 
	\noalign{\smallskip}\hline\noalign{\smallskip}
       $\sigma_\text{shot}$  & 0.1 & 0.1 & 0.1 & 0.1 & 0.2 & 0.3 \\
        \noalign{\smallskip}\hline	
\end{tabular}
\end{center}
\end{table}

Figure~\ref{fig:r0} shows the estimated FWHM of the seeing PSF and the Fried parameter ($r_0$) for the unblocked source observations in Figure~\ref{fig:unblocked_time}. The Fried parameter is an estimate of the strength of the atmospheric turbulence and describes the loss in telescope resolution due to atmospheric blurring, where the resolution will be as if the aperture were decreased to diameter $r_0$. The first thing to notice is how poor the seeing is, typically around 9 arcseconds. This is due to the beam having to traverse a large horizontal path through a thick, turbulent ground layer as it travels from McMath to our observing telescope. The light's proximity to the ground and the rapid changes in the ground elevation results in large changes in the air temperature that generate turbulent cells of varying index of refraction. Assuming the turbulence is dominated by the horizontal path, we estimate the refractive index structure constant to be $C_n^2\sim2\times10^{-13}$.  Scintillation becomes important when the coherence size approaches the Fresnel radius of the turbulence and diffraction from the seeing disk causes interference at the aperture, i.e., when $r_0 \sim\sqrt{\lambda z}$ \cite{Roddier_1981}. With the strong turbulence seen in these observations, the distance from the turbulence at which scintillation dominates is only 400 m, suggesting the scintillation is dominated by the horizontal layer. We attempt to lower systematic errors in the calibration by observing the unblocked source frequently and close in time to the blocked observations to try to sample the same atmospheric conditions.

\subsection{Contrast}
\label{sec:contrast}
We examine the performance of the starshade by calculating the contrast ratio as a function of inner working angle. Following the procedure of Ref.~\cite{Glassman_2014}, we use a signal-to-noise argument to set limits on the contrast from the background noise. We draw photometric subapertures (equal to the size of the PSF) in the image and calculate the standard deviation of pixel counts within that subaperture. This is considered the 1-$\sigma$ noise level in the PSF. Detecting a point source at $n$-$\sigma$ confidence requires the signal to be $n\times\sigma$ higher than the background noise. We define the $n$-$\sigma$ contrast value as the faintest value (normalized by the peak of the unblocked source) for which there is an $n$-$\sigma$ detection, and calculate it as,
\begin{equation}
	C_n = \frac{n\sigma}{F_\ast}
	\label{eq:contrast_calc}
\end{equation}
\noindent where $F_\ast$ is the peak count rate of the unblocked source. 

The dominant source of error in the contrast calculation is the uncertainty in the true intensity of the unblocked source. We calculate lower and upper bounds of the contrast by plugging $F_\ast = \mu_{F_\ast} + \sigma_{F_\ast}$ and $F_\ast = \mu_{F_\ast} - \sigma_{F_\ast}$ into Equation~\ref{eq:contrast_calc}, respectively. Here, $\mu_{F_\ast}$ is the average value and $\sigma_{F_\ast}$ is the 1-$\sigma$ error in the peak count rate of the unblocked source. $\sigma_{F_\ast}$ is calculated by the standard deviation of the ensemble of unblocked images. Section~\ref{sec:unblocked} has shown that this is dominated by scintillation caused by the atmosphere. Increasing the number of exposures should reduce this error, but as Figure~\ref{fig:unblocked_time} shows, there are large systematic offsets that result from sampling different atmospheric conditions. These systematics cannot be reduced by simply increasing the number of images taken. 

The best contrast we achieve at the short baseline is with Vega and is $5.6^{+8.4}_{-2.1}\times10^{-7}$. The best contrast we achieve at the long baseline is with Arcturus and is $2.0^{+1.57}_{-0.49}\times10^{-5}$.

\subsection{Suppression}
\label{sec:suppression}
The contrast ratio is the result of a convolution of many factors external to the starshade. While it is a useful metric to determine the performance of the system, these external factors can dominate over the performance of the starshade alone. A better metric for gauging the starshade performance is suppression. We define suppression as the ratio of the total amount of light entering a telescope when the starshade is in place vs when the starshade is removed. In a space-based starshade mission, the suppression will depend solely on the light suppression capability of the starshade. However, in ground based experiments, external light contributions such as scattering off dust, mirrors, and the atmosphere will limit the ability to distinguish between unsuppressed light and light scattered by external elements. 

The ideal way to calculate suppression is by directly measuring the amount of light incident on the aperture; this can be done with a pupil imaging camera or a simple photodetector. However, for most starshade experiments the only data taken are focal plane images, which have convolved the telescope's response with the light distribution. We attempt to calculate the suppression from the focal plane images by counting all the light that is coming from the edge of the starshade and compare that to the total amount of light from the unblocked source. All of the unsuppressed light will be confined to an annulus (with width equal to the PSF size) around the edge of the starshade. Light from any other location in the image comes from an external source and is not attributed to the performance of the starshade. 

To calculate suppression from the focal plane images, we sum the total number of counts in a box drawn around the starshade that encompasses the starshade diameter plus one PSF width on each side of the starshade. In this way we minimize the contribution from light that is not coming from directly around the starshade. If there is an obvious bright spot on the starshade due to either the starshade stand or a preferred misalignment direction, we mask it out and replace it with the median value around the rest of the starshade. We then sum the total counts in a same-sized box around the unblocked source. Dividing the sum from the blocked images by the sum from the unblocked images gives the suppression value. 

The best suppression we achieve at the short baseline is with Antares and is $4.0^{+73}_{-1.9}\times10^{-4}$. The best suppression we achieve at the long baseline is with Arcturus and is $4.2_{-2.1}\times10^{-4}$, where we do not include an upper limit because the error in the unblocked source brightness was larger than the signal itself. As explained in Section~\ref{sec:photo_contraints}, the suppression calculated at the short baseline is an upper limit to what was actually achieved, and therefore these values should not be taken as representative of the starshade performance. The expected suppression calculated by our optical models are $2\times10^{-5}$ for the short baseline and $6\times10^{-6}$ for the long baseline. However, we expect the observed suppression to be worse at the long baseline as we are dominated by misalignment errors. 

\section{Results}
\label{sec:results}
This section reports the results obtained at the short and long baseline configurations and represents the bulk of the paper. 

\subsection{Short Baseline}
\label{sec:short_baseline}

Observations from the shorter baseline of 570 m are easier for a number of reasons. We are looking through less atmosphere and thus have better image quality, the inner working angle to the starshade is larger and we are less susceptible to misalignment, and we have a wider field of view around the main target that allows us to see background stars. 

In this setup, the best contrast achieved was $5.6\times10^{-7}$ at 30$''$ and we obtained photometric constraints on background stars near Vega. Due to lesser atmospheric extinction (and because this test was initially only a checkout test), the images of the unblocked sources, Vega and Antares, are saturated and we do not have a good photometric calibration. However, we can leverage observations of these targets from the long baseline to provide a calibration, with the caveat that there is significantly more extinction at the long baseline and the contrast and intensity measurements are most likely upper limits. 

Our dominant source of error is variation in the system throughput due to scintillation of the atmosphere. To capture this uncertainty in the unblocked source, we report estimated, upper limit, and lower limit measurements that are calculated using the unblocked value (of mean $\mu_\ast$ and standard deviation $\sigma_\ast$) as either $\mu_\ast$, $\mu_\ast-\sigma_\ast$, or $\mu_\ast+\sigma_\ast$, respectively. Note that this still does not capture the systematic error due to atmospheric extinction.

\subsubsection{Photometric constraints on background stars near Vega}
\label{sec:photo_contraints}
\begin{figure}
	\begin{center}
		\includegraphics[width=\textwidth]{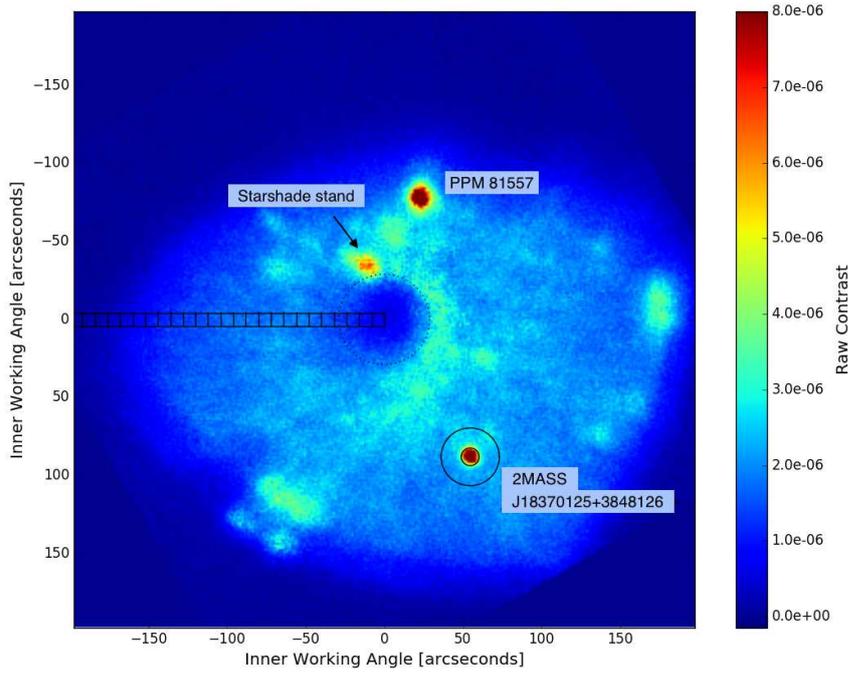}
	\end{center}
	\caption{Stacked image (units of raw contrast) of {\bf Vega} with a total exposure time of 405 seconds from the {\bf short} baseline. The background stars 2MASS J18370125+3848126 and PPM 81557 are labeled. The {\bf dotted circle} in the center corresponds to the IWA. The {\bf boxes} mark the subapertures that the 3$\sigma$ contrast values are calculated in. The {\bf inner} and {\bf outer} circles around J1837 correspond to the photometric and background apertures used to calculate the total signal and total noise, respectively, of the stars.}
\label{fig:vega_img}
\end{figure}

A stacked image of Vega with a total exposure time of 405 seconds is shown in Figure~\ref{fig:vega_img}. These images are shifted to align on the star to the lower right of the starshade. During some of the images, the main heliostat was in motion, so the image of the stars move on the detector, but flaws on the mirror and starshade remain fixed. By correlating the motion of the sources in the image with each other and with the predicted motion of the main source, we can distinguish flaws from astronomical sources. We easily see two companions around Vega, one at the 12 o'clock position and one at the 5 o'clock position. We attribute these to be background stars that are listed in catalogs, but do not have photometric measurements in the visible band. There is also a diffuse scatter around Vega, which appears to be scatter from the mirror. 

\begin{figure}
	\begin{center}
		\includegraphics[width=\textwidth]{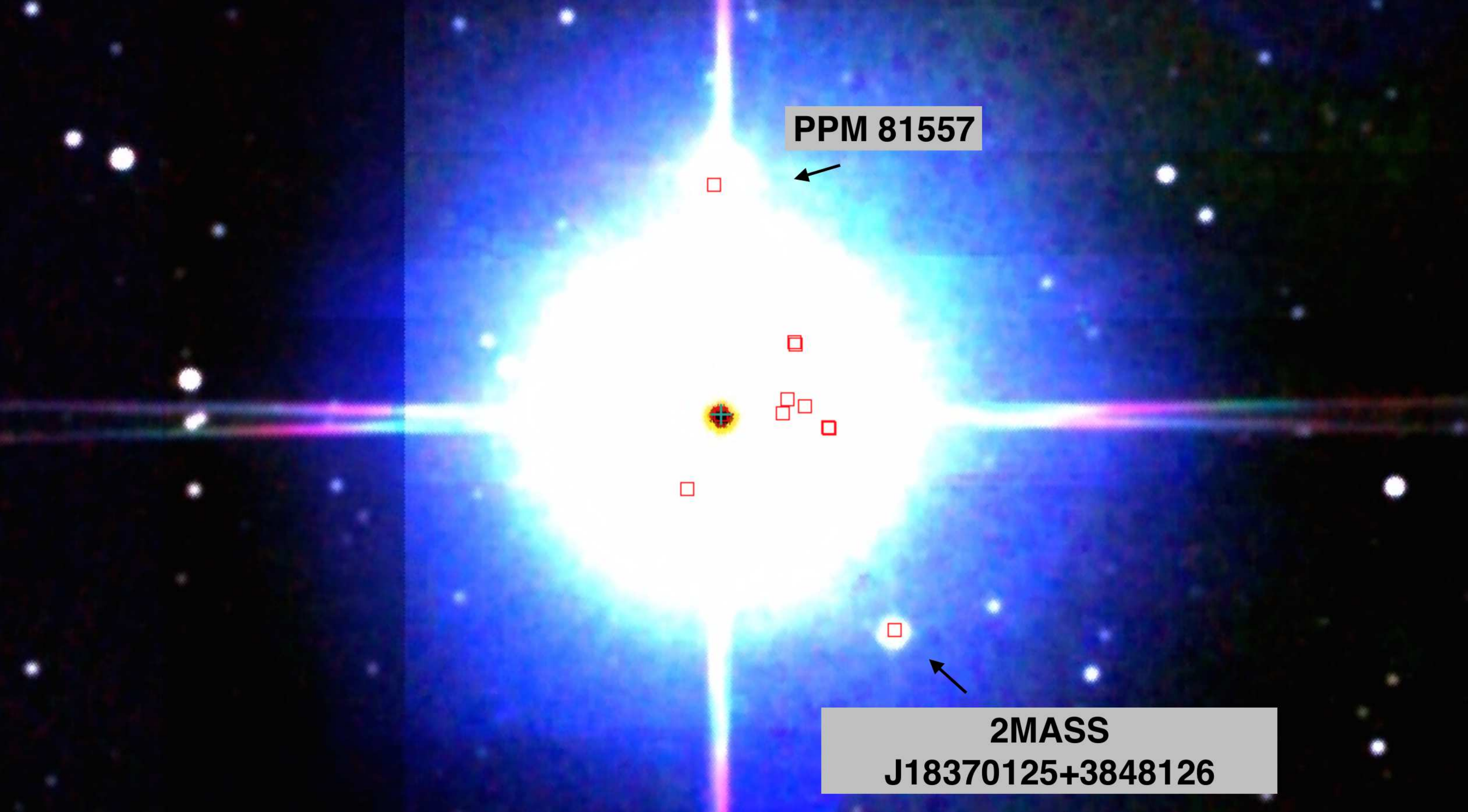}
	\end{center}
	\caption{2MASS image of Vega. The background stars 2MASS J18370125+3848126 and PPM 81557 are labeled. Even at 1.2 $\mu$m, Vega overwhelms the image. PPM 81557 is commonly listed as an artifact in the 2MASS catalog. Image obtained from Ref.~\cite{Simbad}.}
\label{fig:aladin_vega}
\end{figure}

We compare Figure~\ref{fig:vega_img} to a 2MASS image showing the known sources around Vega (Figure~\ref{fig:aladin_vega}) and attribute the two brightest point sources in the image to background stars, 2MASS J18370125+3848126 \cite{2MASS} (hereafter, J1837) and PPM 81557 \cite{PPM}. The measured properties of the background stars are presented in Table~5. We calculate the total signal with a photometric aperture equal to 1.5 times the PSF width (inner circle of black circle in Figure~\ref{fig:vega_img}) and the average background level is calculated using the annulus between the inner and outer (5$\times$ PSF width) circles of Figure~\ref{fig:vega_img}. We calculate the signal-to-noise of a detection using Equation~\ref{eq:snr_vega}, where $S$ is the total number of counts in the photometric aperture minus the average background ($\mu_B$), $n_S$ is the number of pixels in the photometric aperture, $n_B$ is the number of pixels in the background aperture, $RON$ is the readout noise of the detector, and $G$ is the detector gain. We can safely neglect dark noise as the exposures are short and the detector is cooled to -60 $^\circ$C.

\begin{equation}
	SNR = \frac{S}{\sqrt{S + n_s\mu_b\left(1 + \frac{n_s}{n_b}\right) + \left(\frac{RON}{G}\right)^2}}
	\label{eq:snr_vega}
\end{equation}

Since the observations of Vega without the starshade in place are saturated, we must rely on the observations from the longer baseline to serve as the photometric calibration. However, since these observations are at a higher airmass and are taken on a different night, which is sampling different atmospheric conditions, they are not directly transferable to the short baseline observations. The following procedure is used to calibrate the flux of the two background stars. 

\begin{enumerate}
	\item Measure unblocked Vega flux from long baseline
		\begin{itemize}
			\item Discard observations that are above and below 3$\sigma$ from the mean
		\end{itemize}
	\item Integrate Vega spectrum \cite{Bohlin_1996} with detector QE to get expected flux from Vega
		\begin{itemize}
			\item We ignore any other wavelength dependence of the throughput
		\end{itemize}	
	\item The throughput of the system at the long baseline is the measured flux divided by the expected flux
	\item Measure signal and noise from J1837 using the procedure described in the second paragraph of Section~\ref{sec:photo_contraints}
	\item Multiply measured flux by throughput to get theoretical flux of J1837
		\begin{itemize}
			\item This assumes that the atmospheric throughput is the same between the two baselines, which is not true. The throughput for the long baseline will be higher than that of the short baseline, so we will overestimate the intensity of J1837
		\end{itemize}	
	\item Assume a flat spectrum and calculate AB magnitude of J1837
		\begin{itemize}
			\item $m_{AB} = -2.5\log_{10}\left(\frac{f_\nu}{3631 \text{Jy}}\right)$
		\end{itemize}
\end{enumerate}

The measured count rate from unblocked Vega at the long baseline is $(2.5\pm1.5)\times10^{6}~ph/s$. We estimate the expected count rate by integrating the detector response function over a spectrum of Vega \cite{Bohlin_1996}; we neglect any other wavelength dependence on the throughput that could vary between Vega and J1837. The expected count rate is $1.0\times10^{7}~ph/s$, meaning we see a total system throughput (three reflections off mirrors in McMath, atmosphere between McMath and telescope, and telescope optics) of $(24\pm15)$\%, which we believe to be dominated by the atmosphere. Atmospheric extinction on this horizontal baseline should decrease exponentially with decreasing distance, so we expect the atmospheric throughput at the short baseline to be significantly higher and the estimations and error bars should be taken as the worst case scenario.

We use the total system throughput as measured from the unblocked Vega observations to scale the measured count rate of the stars to a theoretical total irradiance in the detector's bandpass (defined where QE $>$ 50\% and assuming a flat spectrum). The estimated spectral irradiance and $AB$ magnitudes of  J1837 and PPM 81557 are presented in Table~5. The quoted error bars are the 1$\sigma$ errors arising from the detection of the source itself and the upper and lower limits are the estimated values if we take the lower or upper limits for the unblocked source value. These limits do not account for the systematic error that is expected due to the unblocked measurements being obtained at a higher airmass. 

\begin{figure}
	\begin{center}
		\includegraphics[width=\textwidth]{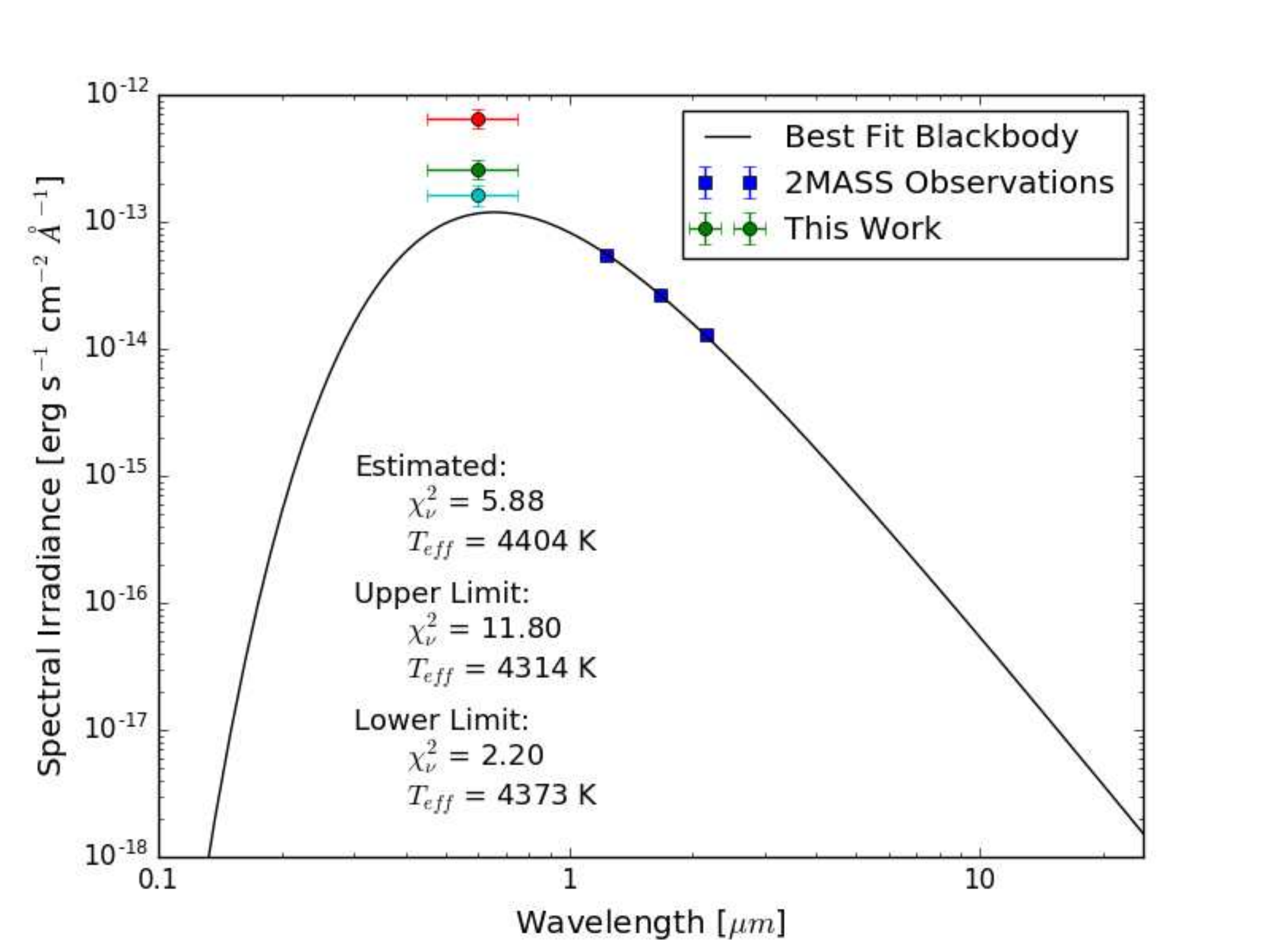}
	\end{center}
	\caption{Best fit blackbody to 2MASS J18370125+3848126 using photometry from 2MASS ({\bf blue squares}) and measurements from this work ({\bf circles}). The different fits correspond to the values calculated using the estimate, upper limit, and lower limit values for the unblocked source. We find it is best fit with the lower limit, which is consistent with our estimate that we are underestimating the value of the unblocked source.}
\label{fig:best_fit_blackbody}
\end{figure}

\begin{figure}
	\begin{center}
		\includegraphics[width=\textwidth]{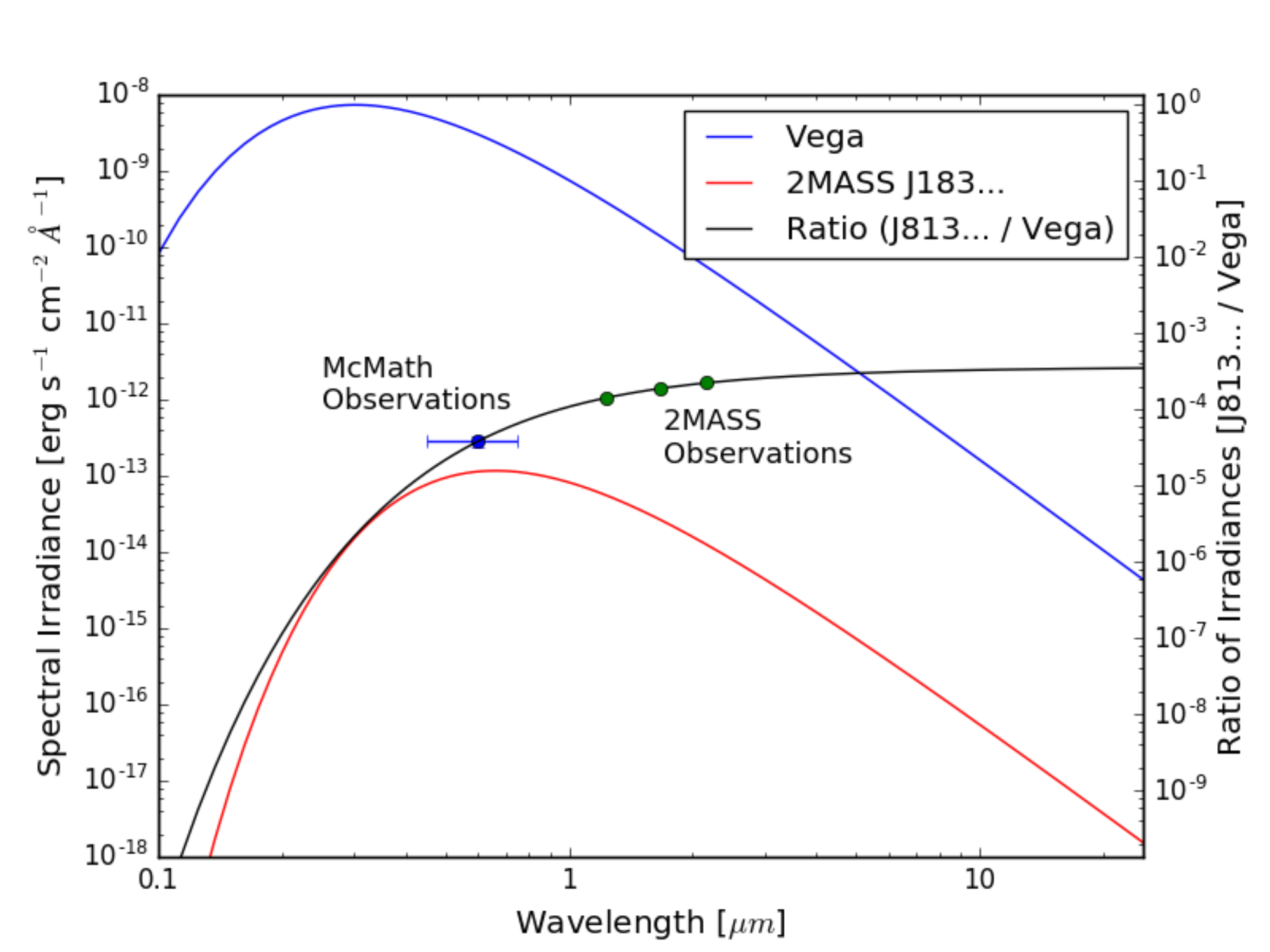}
	\end{center}
	\caption{Model blackbody spectra for Vega ({\bf blue}) and 2MASS J18370125+3848126 ({\bf red}), and the ratio of them ({\bf black}). The two intensities are more similar in the NIR and explains why J1837 has not been measured in the visible.}
\label{fig:star_ratios}
\end{figure}

Combining the observations of J1837 with those from 2MASS \cite{2MASS}, we fit a blackbody to the photometric measurements (Figure~\ref{fig:best_fit_blackbody}). We find the lower limit fits the data best with a reduced chi-squared of 2 and the fit gets better as the photometric data point gets fainter, which is consistent with our expectation of the data points being an upper limit. We find the best fit temperature is 4400 K, suggesting that this star (which we believe to be a K dwarf) is more red than Vega and is why it was previously detected at longer wavelengths only. Figure~\ref{fig:star_ratios} illustrates this point by comparing blackbody curves of the same temperature and intensity as Vega and J1837, and the ratio between the two. In the visible band, J1837 is 10$^{-5}$ times as faint as Vega and rises to 10$^{-4}$ times fainter in the IR, where it was measured by 2MASS \cite{2MASS}. 

PPM 81557 (listed in an astrometric catalogue \cite{PPM}) is brighter than J1837, but is 15 arcseconds closer to Vega. Even in the 2MASS NIR images, it is blended with the PSF of Vega and is flagged as a persistent artifact \cite{Phillips_2010}. Our detection of PPM 81557 adds confidence that this is a real source and not just a 2MASS artifact. We did not find any reported photometric measurements of PPM 81557 in the literature.

\subsubsection{High contrast measurements of Vega}
\begin{figure}
	\begin{center}
		\includegraphics[width=\textwidth]{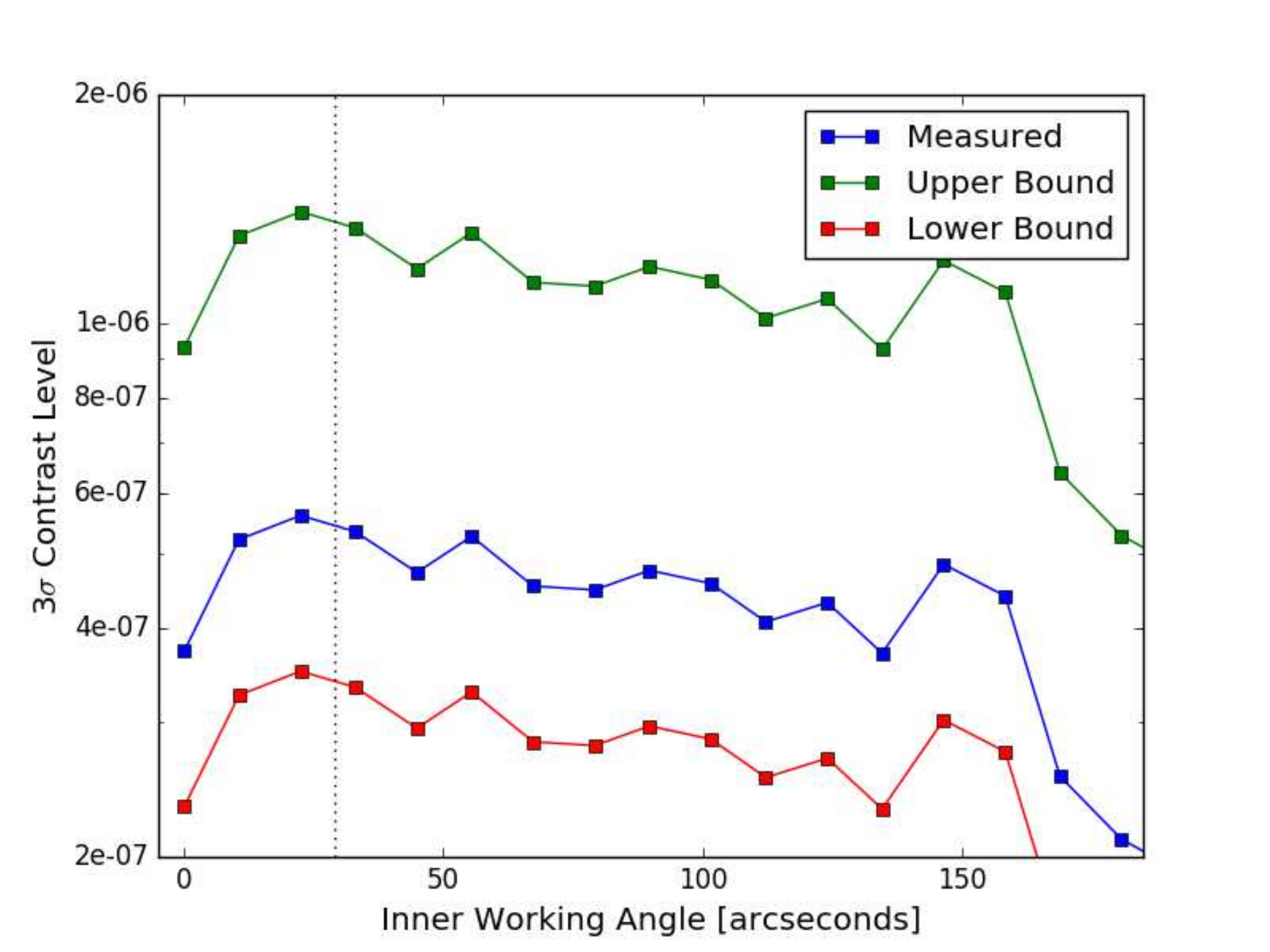}
	\end{center}
	\caption{3$\sigma$ contrast values calculated in the boxes of Figure~\ref{fig:vega_img} for observations of {\bf Vega} at the {\bf short} baseline. The different lines correspond to if the calculations were made with the measured, upper, and lower bound values of the unblocked source. The vertical line denotes the geometric IWA of the starshade.}
\label{fig:vega_3sig}
\end{figure}

We use the unblocked measurements of Vega to convert the images into contrast measurements. Again, since we are underestimating the true intensity of Vega, the contrast achieved is most likely better than reported. We follow the procedure outlined in Section~\ref{sec:contrast} to calculate the 3$\sigma$ contrast level in a series of boxes (seen in Figure~\ref{fig:vega_img}) extending radially from one side of the starshade. These results are shown in Figure~\ref{fig:vega_3sig}, where the measured contrast at the inner working angle of 30$''$ is $5.6^{+8.4}_{-2.1}\times10^{-7}$. Depending on which limit is used for the unblocked source value,  we can set an upper limit on point sources at 30$''$ at $m_V = 15.7$, $m_V = 14.7$, $m_V = 16.2$, for the measured, upper, and lower limits, respectively.

Detecting extended sources is more difficult, particularly with the high level of mirror scatter. As we look at different locations on the sky we sample different sections of the oversized main heliostat mirror and are sampling different scratches, dust, etc. that change the scattered and diffracted light profile. This is clearly seen when comparing the images of Vega (Figure~\ref{fig:vega_img}) and Antares (Figure~\ref{fig:short_antares_img}) from the short baseline. The artifacts in the two images are different, as is the level of diffuse scatter, and it is difficult to determine what is an artifact of the system and what is celestial. In future designs, limiting the number of mirrors in the system and having more stable control and repeatability of the heliostat position will help to mitigate these problems and reference PSF subtractions may be possible to increase signal-to-noise. 

\subsubsection{Possible sources around Antares}
\begin{figure}
	\begin{center}
		\includegraphics[width=\textwidth]{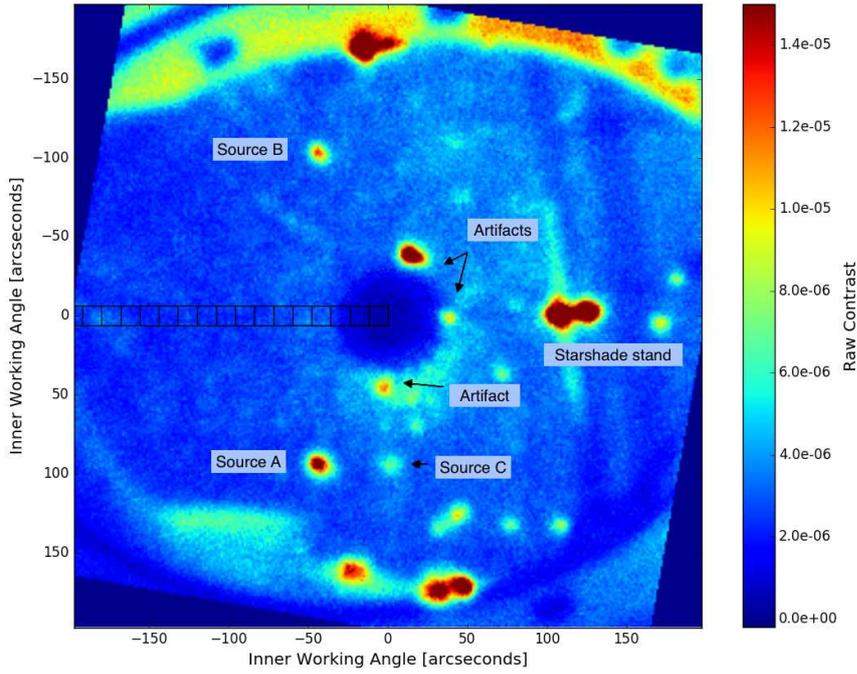}
	\end{center}
	\caption{Stacked image (units of raw contrast) of {\bf Antares} with a total exposure time of 200 seconds from the {\bf short} baseline. The {\bf dotted circle} in the center corresponds to the IWA. The {\bf boxes} mark the subapertures that the 3$\sigma$ contrast values are calculated in. We have labeled obvious artifacts due to the starshade or starshade stand. There are three possible sources that we cannot determine if they are background stars or scratches on the mirror. There are a significant number of fainter point sources in the image as well.}
\label{fig:short_antares_img}
\end{figure}

\begin{figure}
	\begin{center}
		\includegraphics[width=\textwidth]{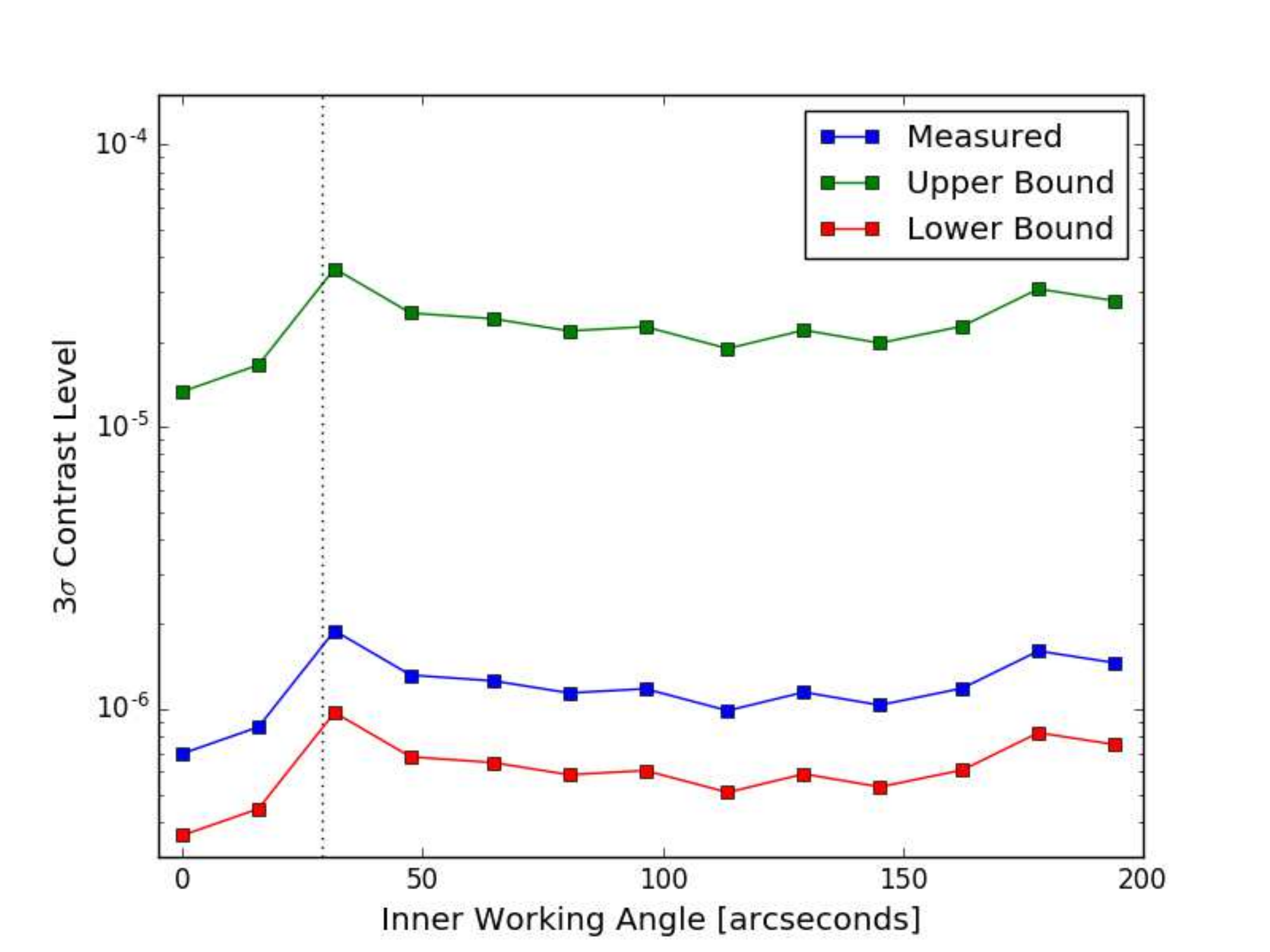}
	\end{center}
	\caption{3$\sigma$ contrast values calculated in the boxes of Figure~\ref{fig:short_antares_img} for observations of {\bf Antares} at the {\bf short} baseline. The different lines correspond to if the calculations were made with the measured, upper, and lower bound values of the unblocked source.}
\label{fig:short_antares_3sig}
\end{figure}

For Antares at the short baseline, we follow the same procedure as the analysis of Vega, again using measurements of unblocked Antares from the long baseline as the calibration source. The combined 200 second image of Antares behind the starshade is shown in Figure~\ref{fig:short_antares_img} and the 3$\sigma$ contrast plot is shown in Figure~\ref{fig:short_antares_3sig}. The measured contrast of Antares at the IWA is $1.8\times10^{-6}$. There are obvious artifacts in the image around the starshade and where the vertical starshade stand crosses a horizontal handrail that is in the beamline. 

There are a number of point sources extended from Antares, but it is not possible to discern if these are background stars or artifacts on one of the mirrors. Unlike the Vega observations, we do not have images where the heliostat is being slewed and cannot correlate the motions of the point sources. We have not found stars in these locations recorded in any star catalogs. The estimated intensity of these sources (if they are celestial) are in Table~5. The horizontal alignment of Source A and Source B and their symmetry around the position of the starshade stand make them suspect as being artifacts of the system.

\begin{table}
    \begin{center}
    \caption{Measured properties of J18370125+3848126 \& PPM 81557 (seen in Figure~\ref{fig:vega_img} around {\bf Vega}) and comparison to their catalog values. The error bars are 1$\sigma$ errors from the detection of the source and the upper and lower limits correspond to which limit of the unblocked source intensity is taken.}
        \begin{tabular}{lcc} 
		\hline\noalign{\smallskip}
		{\bf Source} & {\bf A} & {\bf B}\\ 
            	\noalign{\smallskip}\hline\noalign{\smallskip}
    		{\bf Count rate} [ph/s] & 289.1 & 989.5   \\ 
            	{\bf SNR} & 94.8 & 46.5  \\ 
            	{\bf Measured} $\theta_\text{sep} ['']$ & 92 $\pm$ 15 & 85 $\pm$ 15   \\ 
            	\noalign{\smallskip}\hline\noalign{\smallskip}        
            	{\bf Estimated Intensity}& &\\ 
            	~~~~~[$m_{AB}$] & $10.05\pm0.18$ & $8.71\pm0.07$  \\ 
            	~~~~~[erg s$^{-1}$ cm$^{-2}$ \AA$^{-1}$] & $(2.61\pm0.46)\times10^{-13}$ & $(8.98\pm0.60)\times10^{-13}$  \\ 
            	\noalign{\smallskip}\hline\noalign{\smallskip}
    
            	{\bf Intensity Upper Limit}& &\\ 
           	~~~~~[$m_{AB}$] & $9.06\pm0.18$ & $7.72\pm0.07$  \\ 
           	~~~~~[erg s$^{-1}$ cm$^{-2}$ \AA$^{-1}$] & $(6.53\pm1.14)\times10^{-13}$ & $(22.5\pm1.5)\times10^{-13}$  \\ 
            	\noalign{\smallskip}\hline\noalign{\smallskip}
    
            	{\bf Intensity Lower Limit}& &\\ 
           	 ~~~~~[$m_{AB}$] & $10.57\pm0.18$ & $9.22\pm0.07$  \\ 
           	~~~~~[erg s$^{-1}$ cm$^{-2}$ \AA$^{-1}$] & $(1.63\pm0.29)\times10^{-13}$ & $(5.61\pm0.38)\times10^{-13}$  \\ 
            	\noalign{\smallskip}\hline\noalign{\smallskip}
    
            	{\bf Catalog ID}  & J18370125+3848126 & PPM 81557  \\ 
            	{\bf Catalog} $\theta_\text{sep} ['']$ & 91.69 & 76.49  \\ 
            	{\bf Catalog Intensity} & $m_J$ = 9.402 & $m_B\sim10.8$  \\ 
             	\noalign{\smallskip}\hline	
	\end{tabular}
\end{center}
\label{tbl:vega_companions}
\end{table}
\begin{table}
    \begin{center}
    \caption{Measured properties of the three potential stars seen in Figure~\ref{fig:short_antares_img} around {\bf Antares}. The error bars are 1$\sigma$ errors from the detection of the source and the upper and lower limits correspond to which limit of the unblocked source intensity is taken.}
        \begin{tabular}{lccc} 
            \hline\noalign{\smallskip}
            {\bf Source} & {\bf A} & {\bf B} & {\bf C} \\ 
    	   \noalign{\smallskip}\hline\noalign{\smallskip}
            {\bf Count rate} [ph/s] & 227.5  & 432.3 &  103.9  \\ 
            {\bf SNR} & 24.4 & 45.9 & 11.2  \\ 
            {\bf Measured} $\theta_\text{sep} ['']$ & 100 $\pm$ 15 & 105 $\pm$ 15 & 93 $\pm$ 15  \\ 
    	   \noalign{\smallskip}\hline\noalign{\smallskip}
    
            {\bf Estimated Intensity}& & &\\ 
            ~~~~[$m_{AB}$] & $10.43\pm0.50$ & $9.70\pm0.29$ & $11.27\pm0.88$  \\ 
            ~~~~[$\times10^{-13}$ erg s$^{-1}$ cm$^{-2}$ \AA$^{-1}$] & $1.85\pm1.08$ & $3.60\pm1.10$& $0.85\pm1.07$  \\ 
    	   \noalign{\smallskip}\hline\noalign{\smallskip}

            {\bf Intensity Upper Limit}& & &\\ 
            ~~~~[$m_{AB}$] & $7.22\pm0.50$ & $6.50\pm0.29$ & $8.06\pm0.88$  \\ 
            ~~~~[$\times10^{-13}$ erg s$^{-1}$ cm$^{-2}$ \AA$^{-1}$] & $35.4\pm20.7$ & $69.0\pm20.9$& $16.3\pm20.5$  \\ 
    	   \noalign{\smallskip}\hline\noalign{\smallskip}

            {\bf Intensity Lower Limit}& & &\\ 
            ~~~~[$m_{AB}$] & $11.15\pm0.50$ & $10.43\pm0.29$ & $12.0\pm0.88$  \\ 
            ~~~~[$\times10^{-13}$ erg s$^{-1}$ cm$^{-2}$ \AA$^{-1}$] & $0.95\pm0.55$ & $1.85\pm0.56$& $0.44\pm0.55$  \\ 

	    \noalign{\smallskip}\hline	
        \end{tabular}
\end{center}
\label{tbl:ants_stars}
\end{table}


\subsection{Long Baseline}
\begin{figure}
	\begin{center}
		\includegraphics[width=\textwidth]{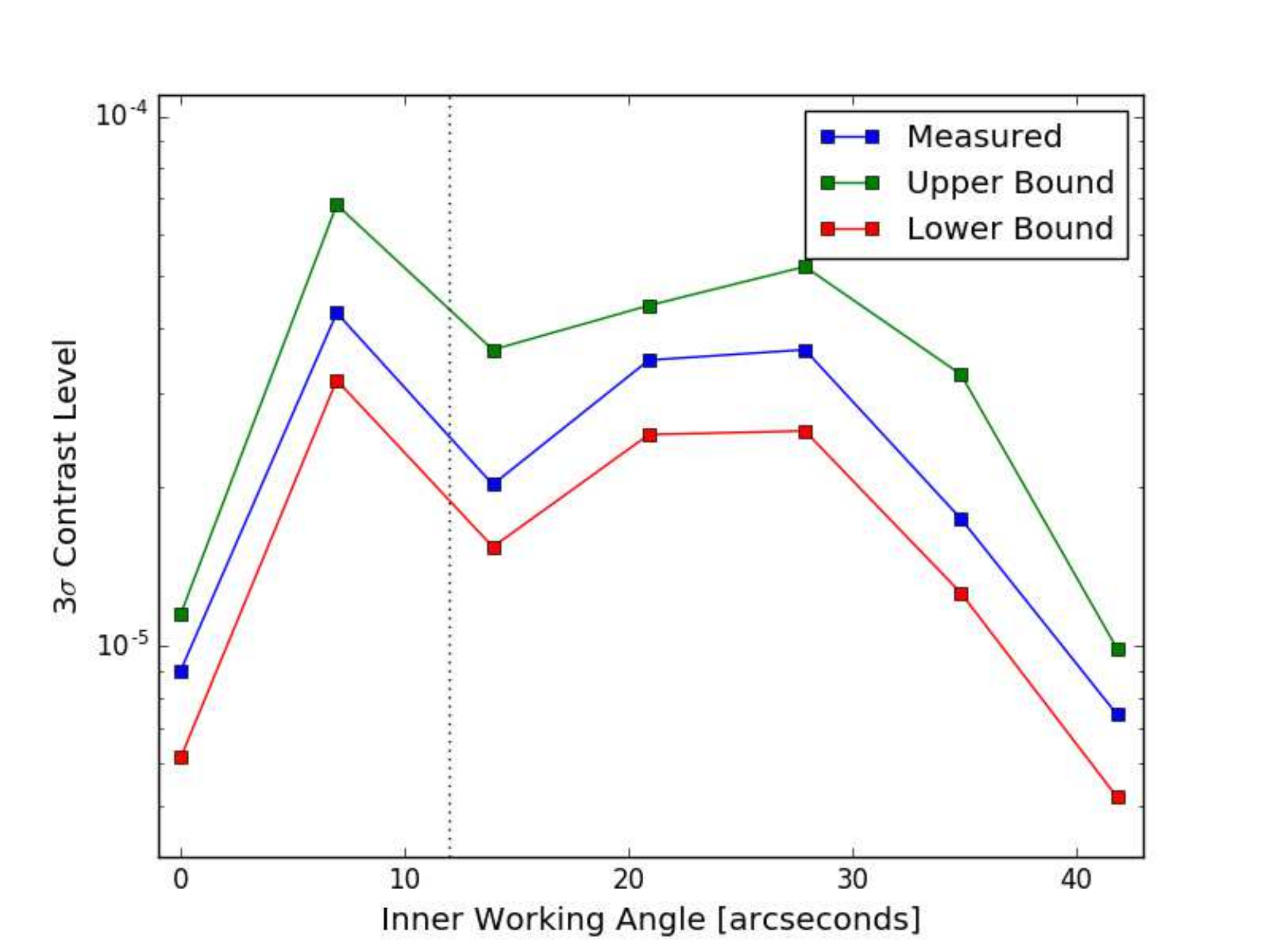}
	\end{center}
	\caption{3$\sigma$ contrast values calculated in the boxes of Figure~\ref{fig:arcturus_img} for observations of {\bf Arcturus} at the {\bf long} baseline. The different lines correspond to if the calculations were made with the measured, upper, and lower bound values of the unblocked source.}
\label{fig:arcturus_3sig}
\end{figure}

Observing from the long baseline brings a host of problems to the observations and data analysis. The problems encountered in Section~\ref{sec:short_baseline} are compounded with a smaller IWA, smaller field of view, and greater atmospheric turbulence. Despite these challenges, we were still able to observe with the most flight-like representation of a starshade to date. The best contrast we achieved from the long baseline at 15 arcseconds IWA was $2.0^{+1.57}_{-0.49}\times10^{-5}$, which was achieved with Arcturus and is presented in Figure~\ref{fig:arcturus_3sig}. 

\begin{figure}
	\begin{center}
		\includegraphics[width=\textwidth]{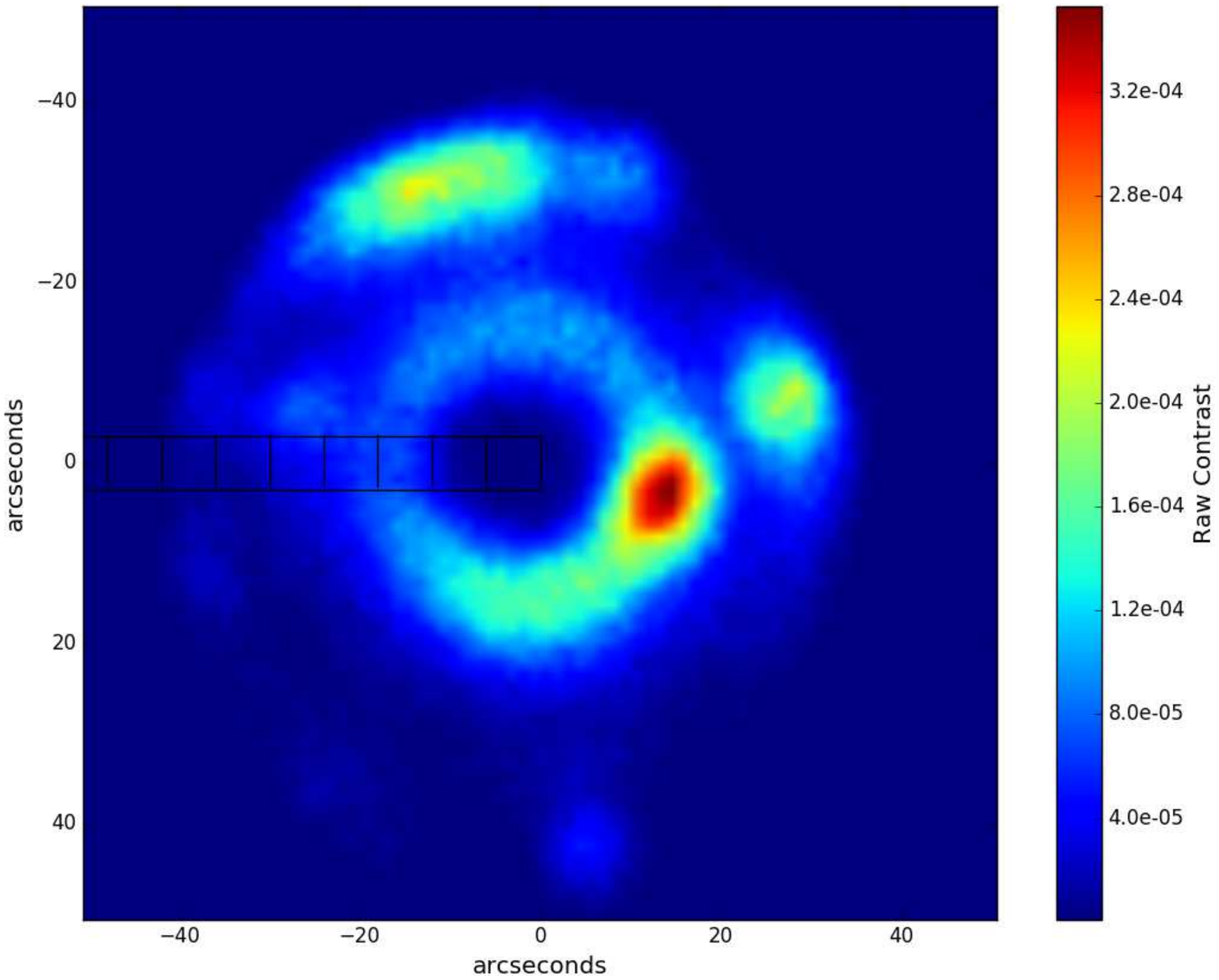}
	\end{center}
	\caption{Stacked image (units of raw contrast) of {\bf Arcturus} with a total exposure time of 177 seconds from the {\bf long} baseline. The {\bf boxes} mark the subapertures that the 3$\sigma$ contrast values are calculated in. We clearly see the ring of light from the tips of the starshade, along with a brighter component where there was preferential misalignment. Diffraction off the edge of the west auxiliary is seen on the outer edge.}
\label{fig:arcturus_img}
\end{figure}

With the large number of short exposures taken, we optimize the image selection to keep those in which we achieve the best contrast and throw out the rest. We convert each image into a raw contrast value, sum the total contrast in the region of interest, and sort the images by that sum. We impose a cut to reject images that have a total contrast above a specified value and median combine to create a master image. There is a tradeoff in the number of images to include in stacking the data, where including more images results in a longer integrated exposure time and will reduce the noise ($\propto\sqrt{t}$ if read noise is negligible). However, as the images are sorted by quality of alignment, including more images will begin to use images of worse alignment where the noise from diffracted light dominates. 

\begin{figure}
	\begin{center}
		\includegraphics[width=\textwidth]{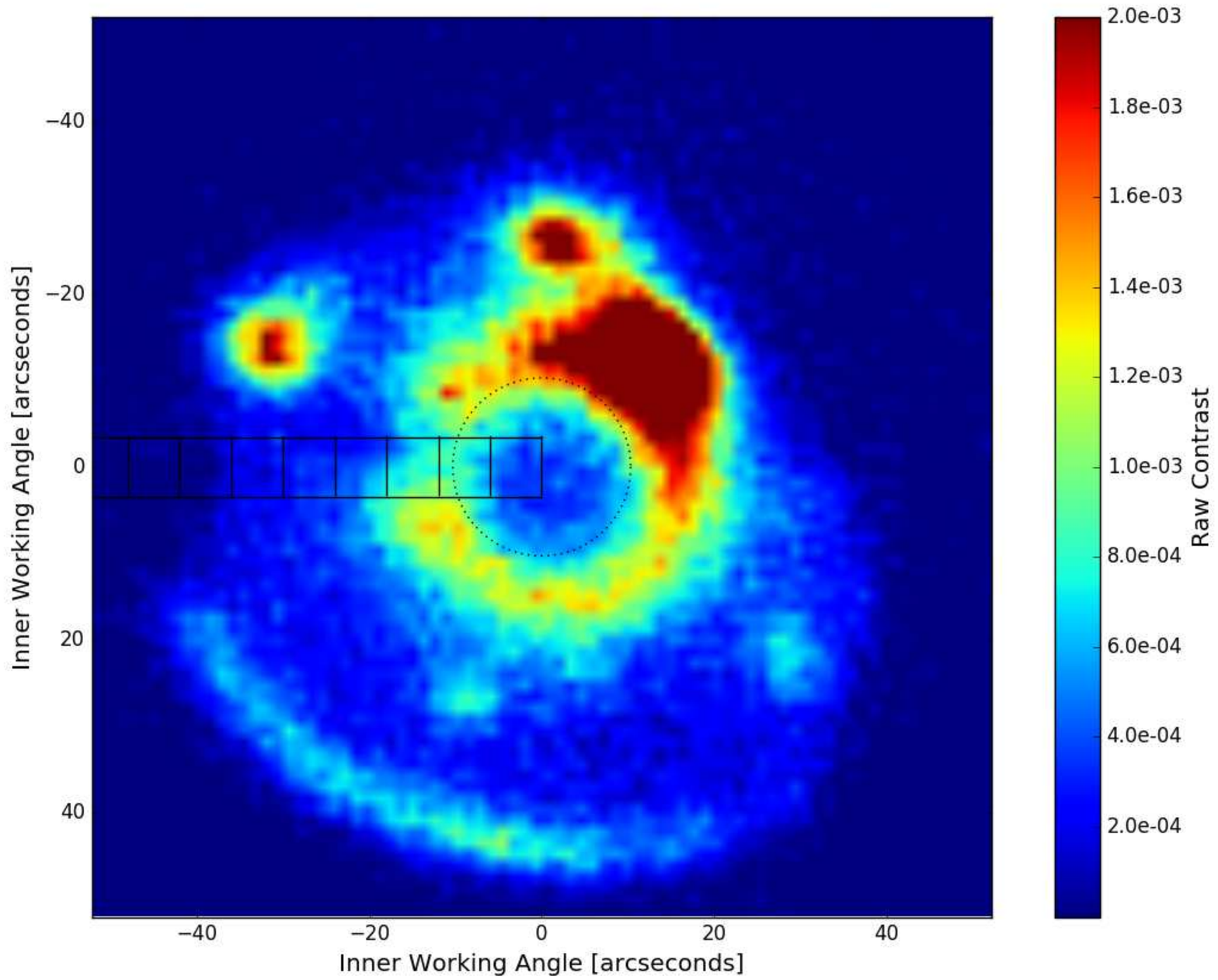}
	\end{center}
	\caption{Stacked image (units of raw contrast) of {\bf Fomalhaut} with a total exposure time of 120 seconds from the {\bf long} baseline.}
\label{fig:fomalhaut_img}
\end{figure}

\begin{figure}
	\begin{center}
		\includegraphics[width=\textwidth]{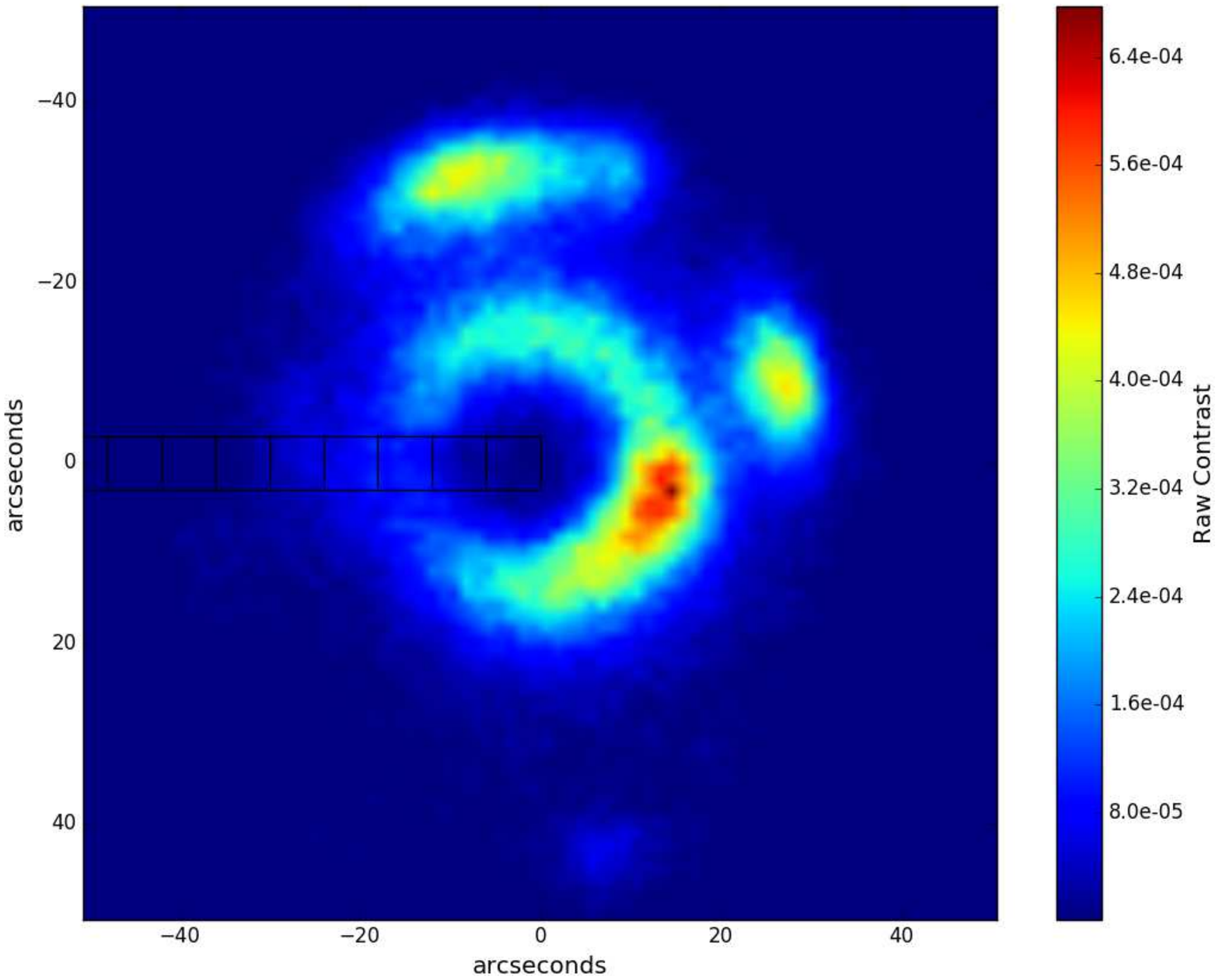}
	\end{center}
	\caption{Stacked image (units of raw contrast) of {\bf Vega} with a total exposure time of 40 seconds from the {\bf long} baseline.}
\label{fig:long_vega_img}
\end{figure}

The arc of light at the edge of the west auxiliary mirror that is seen in the long baseline images (Figures~\ref{fig:arcturus_img}, \ref{fig:long_vega_img}) is light diffracted by the starshade 140 m upstream, that is again diffracted as the west auxiliary mirror clips the beam. In Figure~\ref{fig:fomalhaut_img}, the starshade was moved so that it was only 5 m upstream of the west auxiliary and therefore has less distance for the diffraction to spread to a larger angle before being clipped by the west auxiliary. In this configuration, the arc is still present, but to a lesser extent. Future use of this facility should minimize the number of mirrors in the system and apodize the outer edges of the heliostat mirrors to limit diffraction from the mirrors.

Figures~\ref{fig:arcturus_img} - \ref{fig:long_vega_img} contain the results from long baseline observations of Arcturus, Fomalhaut, and Vega.

\subsection{Formation Flying}
The large dataset of images obtained with the starshade in many different configurations and alignments provide an extensive source of information for future studies of formation flying with starshades, a technology area in need of development for future missions. The tracking system \cite{Harness_2016} that maintained alignment of the star and starshade is essentially ``formation flying'' and uses algorithms similar to what will be used by future formation flying sensors. A future direction of this project is to integrate data from the science camera into the feedback loop, using the distribution and intensity of light diffracted around the starshade as the guiding signal. This method of long range position sensing has been identified as a potential method to be used in future starshade missions \cite{Scharf_2016}. With this large dataset in hand, we can begin designing and training the algorithms that turn focal plane images of light diffracting around a starshade into control signals for formation flying. The analysis of the data in this context is deferred to a later publication.

\section{Discussion}
\label{sec:discussion}
This experiment was the first demonstration of a new type of facility, using a heliostat to maintain alignment of astronomical sources with a starshade. As is expected with most demonstrations of a new technology, we did not achieve the full capability of the system, but were able to identify the viability of the facility and the limiting factors. The key question to be answered by the on-sky tests was: ``can a heliostat be adapted to do astronomy with starshades?''. We believe the work presented here has answered that question in the affirmative. Answering the second-order question of how useful this facility is for doing cutting edge science and technology development of starshades depends on the ability to overcome the identified limitations of the system. 

Using optical models to calculate the predicted performance of the starshade, we can quantify the achieved performance with respect to external factors. In the long baseline configuration, the best contrast achieved at 15 arcseconds was $2\times10^{-5}$. Our models predict that contrast should be $2\times10^{-7}$. The model that includes 
wavefront errors from an atmosphere with 8 arcsecond seeing incident on the starshade predicts the contrast to be $2\times10^{-6}$, bringing the model closer to the observations. The remaining discrepancy between observations and model can be attributed to misalignment in the starshade. Because our tracking system was not able to correct for misalignments on short timescales, most of the images were obtained while the starshade was slightly misaligned. Our models show the contrast will be 10$\times$ higher if we are offset by only 3 cm.  

In the short baseline configuration, the best contrast achieved at 30 arcseconds was $6\times10^{-7}$. Our models predict that contrast should be $2\times10^{-7}$. As mentioned previously, the atmospheric effects from the short baseline are significantly less than that of the long baseline. We are instead dominated by scatter off the heliostat mirrors. Calculated from the diffuse light in Figure~\ref{fig:vega_img}, scatter can easily account for the factor of 3 discrepancy between the models and observations.

We would like to note that the starshades we tested were limited to $10^{-6}$ suppression and 10$^{-7}$ contrast by manufacturing of the tips of the starshade petals. For these initial tests, we did not focus on manufacturing the starshades to a tolerance that would enable the full $10^{-10}$ contrast that is possible, but instead focused on a durable starshade that could be made cheaply. Once we improve the siderostat system to minimize the effect of external factors, we can manufacture the starshade to the tolerance needed to achieve a flight-like contrast level.

By far the dominant limitation of this system is the atmosphere. Broadening of the PSF due to atmospheric seeing worsens the contrast at the starshade's IWA and reduces the search space for astronomical targets. Scintillation in the atmosphere induces variance in the true value of the source and leads to an uncertainty in interpreting results in terms of starshade performance. Moving to a site at a higher elevation and with less atmosphere will significantly improve the usefulness of the system. 

The next largest limitations in the system are the tracking system of McMath's aged system and scatter off dust in the multiple mirrors in the system. It is beyond the scope of this paper to provided detailed numbers on what is needed to reach the full potential of using starshades on the ground with heliostats. We do note however, that most elements of this system were far from ideal and slight improvements to any aspect will lead to large gains in capability. The following section summarizes what can be done in the future to improve the capability of this concept. 

\section{Improvements for Future Tests}
\label{sec:improvements}
We briefly summarize the lessons learned from these observations and how this work can inform future observations using siderostats and starshades.
\begin{itemize}
	\item[$\bullet$] Misalignment of the starshade is the limiting factor in achieving high contrast
	\begin{itemize}
		\item[--] Move to a site with less atmosphere and more laminar wind flows
		\item[--] Have faster control on the position of the starshade
		\begin{itemize}
			\item[*] This could be done with better tip/tilt control of the siderostat
			\item[*] Or, use the siderostat for rough alignment and put the starshade on a translation stage for quick and precise alignment
		\end{itemize}
	\end{itemize} 
	\item[$\bullet$] Variance in the unblocked source intensity leads to large errors in contrast measurements
	\begin{itemize}
		\item[--] More frequent observations of the unblocked source need to be made and closer in time to the blocked observations
		\item[--] A larger number of unblocked images need to be made to knock down the intensity sample variance
		\item[--] Using neutral density filters will allow for exposure times comparable to those of the blocked observations to sample the same scintillation statistics
		\item[--] A calibration source that has a constant and known intensity could be used to have a relative measurement of the scintillation for every image.
	\end{itemize}
	\item[$\bullet$] Atmospheric turbulence is ruining image quality, causing variance in intensity, and lowering the throughput of the system
	\begin{itemize}
		\item[--] Move to a site with less atmosphere
		\item[--] Use adaptive optics with observing telescope
	\end{itemize} 
	\item[$\bullet$] Scatter and diffraction from the mirrors will limit contrast after the misalignment problem has been solved 
    	\begin{itemize}
    		\item[--] Use cleaner, higher quality mirrors
		\item[--] Use fewer mirrors in the system
		\item[--] Apodize the outer edge of the mirrors
    	\end{itemize} 	
	\item[$\bullet$] The movement of the siderostat to compensate for tracking errors is constantly moving the star field and makes it difficult for image alignment
	\begin{itemize}
		\item[--] Use a laser guide star to have an off-axis field star at all times for image alignment and photometric calibration
	\end{itemize} 
\end{itemize}

\section{Conclusions}
\label{sec:conclusions}
We have demonstrated a practical method to doing astronomy with starshades, achieved high contrast ($5.6\times10^{-7}$) at a moderate IWA (30$''$) on astronomical targets, and provided photometric measurements of stars near Vega in a new wavelength regime. These stars were previously inaccessible in visible light due to their proximity to Vega. The systematic errors in the measurements are large due to the loose constraint on the intensity of the unblocked source, but still provide useful constraints. This was the first demonstration of the only system capable of doing astronomy with starshades and testing macroscopic starshades at the correct Fresnel number. 

At the long baseline, the contrast  was limited by constant misalignment from errors in McMath's tracking system and disturbances in the atmosphere, but we have identified the practicalities in using a starshade-heliostat system for astronomy. Section~\ref{sec:improvements} outlines what can be done to improve future observations. 

The contrast we were able to achieve at the short baseline is impressive, given the conditions and quality of optics we were working with, and speaks volumes about the durability, efficiency, and potential that starshades possess. While an IWA of 30'' seems rather large, we try to place these observations in perspective by noting that we achieved high contrast ($<6\times10^{-7}$) with broadband visible light, at an IWA of 4 $\lambda/D$, with a 5 minute exposure and a 2 cm aperture. 

We have taken the first steps in understanding astronomical observations with a starshade. The large dataset obtained, with the starshade in many different configurations and alignments, provide a useful training set for extracting signals out of the noise of starshade images. Most importantly, we have demonstrated the practicality of astronomical observations with a starshade and have taken the first steps towards building a rich experience base with starshades. 

In a final note, our efforts to test siderostats have been on hold since NSF funding for McMath expired in late 2016 and the facility is being shut down. As of the writing of this paper we are actively seeking support to return to McMath to take the observations and the understanding of starshade systems to the next level. We also hope to help keep the old site open and available to the astronomy community as an instrument test and development facility instead of simply closing it forever.

\begin{acknowledgements} 
This work was supported by a Strategic University Research Partnership between CU and JPL with Co-I Randy Pollock. AH was supported at CU by a NASA Space Technology Research Fellowship (NNX13AM71H). The authors would like to thank Richard Capps (JPL) for suggesting the use of McMath. The authors would like to thank Detrick Branston for excellent observing support at McMath and Ann Shipley, Ben Zeiger, Danny Smith, and Michael Richards for assistance during observing runs. AH would also like to thank Wayne Green. The authors thank the anonymous referee for the very useful suggestions and comments. The McMath-Pierce Solar Telescope facility is operated by the National Solar Observatory. The National Solar Observatory is operated by the Association of Universities for Research in Astronomy under a cooperative agreement with the National Science Foundation.
\end{acknowledgements}

\bibliographystyle{spmpsci}       
\bibliography{exastro_paper.bib}   

\end{document}